\documentclass[journal]{IEEEtran}
\usepackage{amsmath}
\usepackage{amssymb}
\usepackage{amsthm}
\usepackage{cite}
\usepackage{color}
\usepackage{xcolor}
\usepackage{caption}
\usepackage{epsfig,latexsym}
\usepackage{float}
\usepackage{fancyhdr}
\usepackage{graphicx}
\usepackage{indentfirst}
\usepackage{mdwmath}
\usepackage{mdwtab}
\usepackage{subfigure}
\usepackage{setspace}
\usepackage{times}
\usepackage{url}
\usepackage{multirow}
\usepackage{algorithm,algorithmic}

\newtheorem{remark}{Remark}
\newtheorem{theorem}{Theorem}

\newtheorem{lemma}{Lemma}

\newtheorem{corollary}{Corollary}

\newtheorem{proposition}{Proposition}

\begin{document}

\title{On the Performance of Uplink Pinching Antenna Systems (PASS)}
\author{Tianwei Hou,~\IEEEmembership{Member,~IEEE,}
        Yuanwei Liu,~\IEEEmembership{Fellow,~IEEE,}
        Arumugam Nallanathan,~\IEEEmembership{Fellow,~IEEE}

\thanks{Tianwei Hou is with the School of Electronic and Information Engineering, Beijing Jiaotong University, Beijing 100044, China, and also with the School of Electronic Engineering and Computer Science, Queen Mary University of London, London E1 4NS, U.K. (email: twhou@bjtu.edu.cn).}
\thanks{Yuanwei Liu is with the Department of Electrical and Electronic Engineering, The University of Hong Kong, Hong Kong (e-mail: yuanwei@hku.hk).}
\thanks{Arumugam Nallanathan is with the School of Electronic Engineering and Computer Science, Queen Mary University of London, London E1 4NS, U.K., and also with the Department of Electronic Engineering, Kyung Hee University, Yongin-si, Gyeonggi-do 17104, Korea. (e-mail: a.nallanathan@qmul.ac.uk).}

}

\maketitle
\vspace{-0.3in}
\begin{abstract}
Pinching antenna (PA) is a flexible antenna composed of a waveguide and multiple dielectric particles, which is capable of reconfiguring wireless channels intelligently in line-of-sight links. By leveraging the unique features of PAs, we exploit the uplink (UL) transmission in pinching antenna systems (PASS). To comprehensively evaluate the performance gains of PASS in UL transmissions, three scenarios, multiple PAs for a single user (MPSU), a single PA for a single user (SPSU), and a single PA for multiple users (SPMU) are considered. The positions of PAs are optimized to obtain the maximal channel gains in the considered scenarios. For the MPSU and SPSU scenarios, by applying the optimized position of PAs, closed-form expressions for analytical, asymptotic and approximated ergodic rate are derived. As the further advance, closed-form expressions of approximated ergodic rate is derived when a single PA is fixed in the SPMU scenario. Our results demonstrate the following key insights: i) The proposed PASS significantly outperforms conventional Multiple-input Single-output networks by exploiting the flexibility of PAs; ii) The PA distribution follows an asymmetric non-uniform distribution in the MPSU scenario; iii) Optimizing PA positions significantly enhances the ergodic sum rate performance.
\end{abstract}

\begin{IEEEkeywords}
Line-of-sight, performance analyse, pinching antenna, PASS, uplink transmission.
\end{IEEEkeywords}

\vspace{-0.1in}
\section{Introduction}

With the increasing demand for flexibility in next generation wireless networks, the conventional fixed antennas cannot fully satisfy the growing requirements.
Multiple antenna technique has emerged as a cornerstone in advancing wireless communication systems~\cite{MIMO_1_Mag, MIMO_Mag_2}. Over the evolution from the first generation to the sixth generation (6G) of wireless communication systems, antenna pattern design has undergone rapid evolution, significantly enhancing channel gains in wireless communications. For example, in 6G, the extra-large-scale massive multiple-input multiple-output (MIMO) technology has attracted substantial attention due to its ability to deploy massive discrete antennas along the edges of room ceilings, thereby achieving a higher degree of freedom~\cite{EL-MIMO_3}. However, fixed antenna systems require a substantial number of antennas, making them less cost- and energy-efficient.

In next-generation (NG) wireless networks, traditional antenna technologies fall short of addressing the growing demands of diverse applications~\cite{NG_network_Mag_8-1,NG_network_Mag_8-2}. As a result, innovative antenna patterns have become both necessary and expected. One promising solution lies in utilizing novel meta-materials to enable reconfigurable intelligent surface (RIS) technology, which allows precise control over the reflected electromagnetic (EM) waves~\cite{RIS_mag_4,RIS_Hou_5}. By doing so, the small-scale channel gain between the base station (BS) and users can be intelligently coherent in a more manner way. Additionally, advanced materials enable the antenna aperture to function as a multi-layered structure, where EM waves propagate and interact within a three-dimensional aperture, which paves the way for significantly enhanced communication capabilities, particularly in scenarios requiring high spectral efficiency and flexible spatial resource management~\cite{HMIMO_7,Holo_MIMO_18,huang2024holographicintegrateddataenergy_19}. These development significantly boosts communication capabilities, particularly in applications demanding high spectral efficiency and flexible spatial resource management.

Building on the aforementioned contributions, the observed performance gain primarily hinges on small-scale fading channels. Therefore, one natural question raises: can performance be further enhanced by optimizing the large-scale fading channels? Fortunately, several emerging techniques present viable solutions, including unmanned aerial vehicle (UAV) communications~\cite{UAV_Mag2}, movable antenna systems~\cite{Movable_Antenn_Mag}, and fluid antenna systems~\cite{fluid_anntena_mag, fluid_anntena_mag2}. From the perspective of spatial coverage, UAV communication systems offer the capability to optimize large-scale fading over a kilometer-scale range~\cite{UAV_paper1}. The performance of both single UAV networks and cellular UAV networks has been thoroughly evaluated in~\cite{UAV_Hou_2}, demonstrating that UAVs can significantly reduce the path loss exponent by intelligently maneuvering into line-of-sight (LoS) environments, thereby improving channel gains. Additionally, the flexible mobility of UAV platforms enables BSs deployed on UAVs to move to optimal locations for serving ground users through trajectory optimization~\cite{UAV_traj_opt1}. However, due to the physical size and operational constraints of UAVs, UAV communications are best suited for outdoor scenarios, which is not practical for indoor scenarios, where alternative approaches are required and expected to overcome communication challenges.

The focus then shifts to centimeter-level. In NG networks, the typical antenna aperture spans meters and comprises thousands of antenna elements. Over time, the limitations of MIMO have become apparent. A large-scale antenna array comes with huge cost of expensive radio frequency (RF) chains and enormous power consumption. To address these challenges, the concept of fluid antennas (FA) has been proposed, advocating for shape and position flexibility~\cite{Fluid_one}.
With the aid of FA, antennas can dynamically reposition themselves through a network of ports, optimizing their location to enhance both large-scale and small-scale fading channels. The adaptability of FA enables antenna positioning within the centimeter-to-meter range~\cite{Fluid_antenna_paper1}. Moreover, the spatial modeling of FA systems was studied in~\cite{Fluid_new_spacial}, highlighting that performance gains primarily stem from improvements in small-scale fading channels. However, it is noteworthy that in LoS channels, the performance advantages of FA systems over conventional fixed-antenna systems diminish, demonstrating a limitation of FA in such indoor communication scenarios.

The aforementioned techniques have limited applicability or performance gains in indoor scenarios, which motivates the exploration of pinching antenna (PA) systems (PASS)~\cite{liu2025pinchingantennasystemspass,yang2025pinchingantennasprinciplesapplications}. The primary components of PASS include a waveguide and multiple dielectric particles, such as plastic pinches. In PASS, the waveguide, connected to the access point (AP), is positioned along the edge of the ceiling. The waveguide is considered isolated from the indoor free space, with EM waves propagating through waveguide~\cite{waveguide_mag}. Multiple PAs are deployed along the waveguide, enabling the EM waves to radiate or be received through these PAs~\cite{PAN_1,example2025}. In~\cite{ding2024flexibleantennasystemspinchingantennaperspective}, the potential of PASS for indoor wireless communication was introduced, incorporating both orthogonal multiple access (OMA) and non-orthogonal multiple access (NOMA) techniques for downlink transmissions. The performance of multiple PAs deployed on a single waveguide and on multiple waveguides under LoS channels was analyzed in closed-form. Building on this, a dynamic PA selection approach for NOMA transmission was proposed by utilizing a spherical wave channel model in~\cite{wang2024antennaactivationnomaassisted}. The array gain of multiple PAs in a single waveguide was analysed in two cases, fixed inter-antenna spacing with PAs and fixed number of PAs with fixed inter-antenna spacing~\cite{ouyang2025arraygainpinchingantennasystems}. Furthermore, the uplink (UL) data rate of PASS in OMA transmission was maximized by employing successive convex approximation to optimize the allocation of time resource blocks in~\cite{tegos2024minimumdataratemaximization}.

\subsection{Motivation and Contribution}

From the above discussion, it is clear that existing studies have primarily focused on the downlink transmission of PASS, with limited attention dedicated to uplink (UL) transmission. Although UL transmission has been partially addressed in~\cite{tegos2024minimumdataratemaximization}, a comprehensive performance analysis of PASS in UL remains largely unexplored. The main motivations for this work are summarized as follows:
\begin{itemize}
    \item Since the waveguide is a one-dimensional material capable only of delaying signals, deploying multiple PAs for a single user in UL transmission may offer a more practical solution than downlink transmission.
    \item Key questions remain unanswered: How many PAs are optimal for users? Should multiple PAs be deployed for a single user, or should a single PA serve multiple users?
    \item The optimized spatial distribution of PAs in UL transmission has not yet been determined.
    \item An in-depth evaluation into the ergodic rate performance of PASS in UL transmission is needed, particularly when the number of PAs is sufficient or limited.
\end{itemize}

To address these gaps, this article aims to tackle the aforementioned challenges by considering varying numbers of PAs, thereby evaluating and optimizing the performance of PASS in UL transmission.

The following is a brief summary of our main theoretical contributions with regard to the UL PASS set-up in this article:
\begin{itemize}
 \item We exploit a PASS for UL LoS transmissions, where multiple PAs are deployed along a single waveguide to serve multiple users. In order to provide a comprehensive analysis, we consider three distinct scenarios: Using multiple PAs for a single user (MPSU), using a single PA for a single user (SPSU) and using a single PA for multiple users (SPMU).
 \item For the MPSU scenario, we begin by deriving the optimized position distribution of multiple PAs in the near-zone (NZ) and far-zone (FZ) schemes. Subsequently, we derive closed-form expressions for the ergodic rate of individual users. To provide deeper insights into the system performance, we further obtain analytical and asymptotic results for the ergodic rate performance. Our analysis reveals that the optimized locations of PAs exhibit an asymmetric non-uniform distribution in the NZ scheme, whereas the optimized locations of PAs approach symmetric uniform distribution in the FZ scheme.
 \item For the SPSU scenario, we derive closed-form expressions for the ergodic rate of individual users when a single PA is associated to a single user. Additionally, the analytical, asymptotic results and approximated results are evaluated in the high-signal-to-noise (SNR) regime. Our analysis demonstrates that the deployment of multiple PAs significantly enhances the ergodic rate performance between each PA and its associated user, which highlights the distinctive performance advantages of utilizing PASS in wireless communications.
  \item To further investigate the performance of PASS in the SPMU scenario, the ergodic rate is evaluated under a limited number of PAs, considering both fixed PA and optimized PA cases. In the fixed PA case, a single PA is positioned at the center of the waveguide to serve multiple randomly deployed users, with exact and approximated results derived accordingly. For the optimized PA case, the optimal PA position is obtained in closed-form, revealing that in UL transmissions, the optimal PA placement is influenced by the distance ratio among multiple users.
  \item Our simulation results validate our analysis, demonstrating that: 1) The proposed PASS outperforms conventional networks by effectively mitigating large-scale fading channels, showcasing its unique advantage; 2) In the MPSU scenario, by adopting multiple well designed PAs, the antenna gain can be approximated to the number of PAs; and 3) The ergodic sum rate in the MPSU scenario exceeds that of the SPSU and SPMU scenarios, which indicates that optimizing the PA position is strongly recommended for enhancing the ergodic sum rate performance.
  \end{itemize}

\subsection{Organization and Notations}

The remainder of this article is organized as follows. Section \uppercase\expandafter{\romannumeral2} discusses the model of PASS, where the spatial model in LoS channels are demonstrated. Our analytical results of the PASS are presented in Section \uppercase\expandafter{\romannumeral3}, where the MPSU, SPSU and SPMU scenarios are separated in three subsections. Section \uppercase\expandafter{\romannumeral4} provides the numerical results of our proposed PASS. In Section \uppercase\expandafter{\romannumeral5}, we conclude our article. $\mathbb{E}(\cdot)$ denotes the expectation.

\section{System Model}

\subsection{Antenna and Channel Model}
\begin{figure}[t!]
\centering
\includegraphics[width =2.5 in]{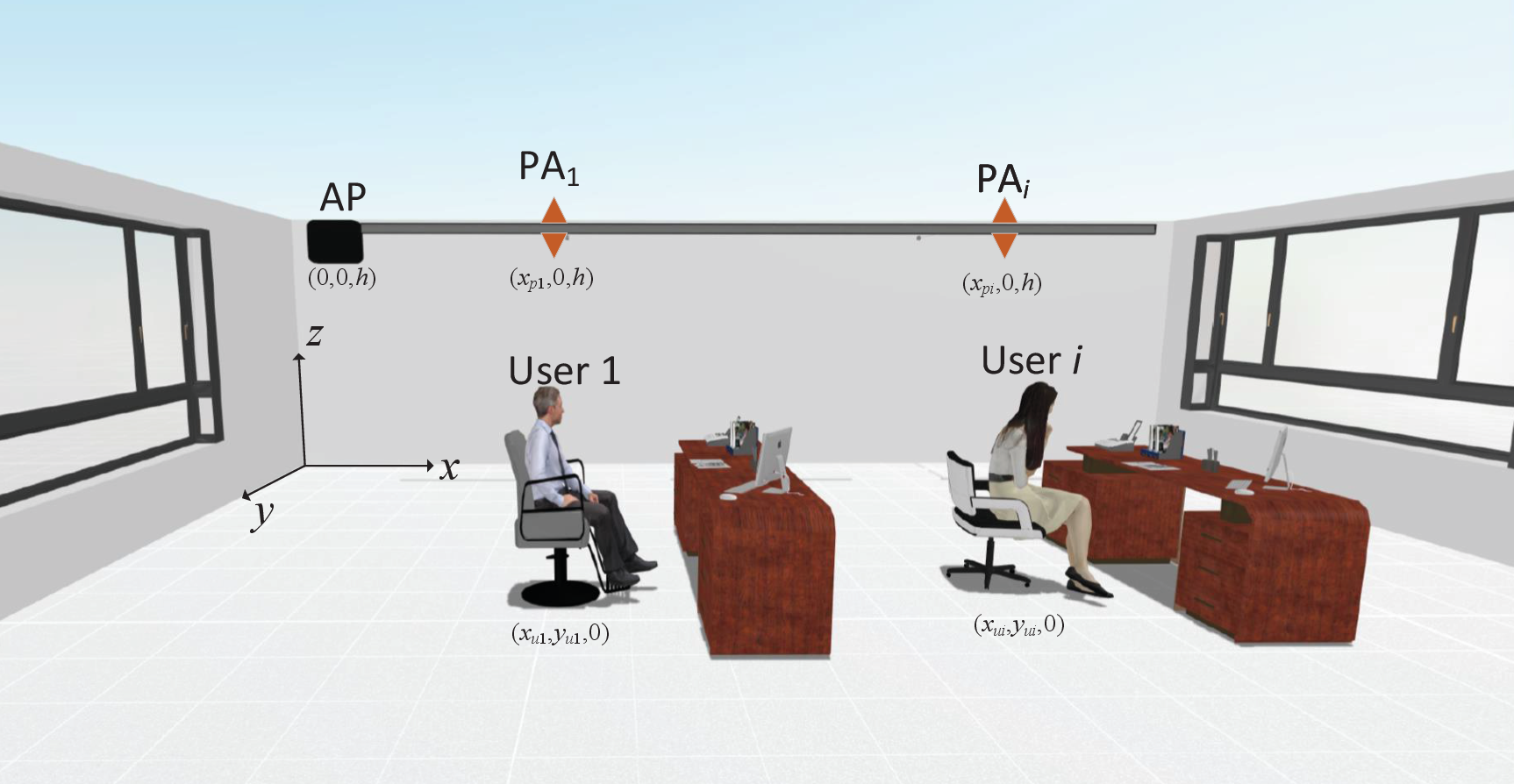}
\caption{Illustration of the PASS, where multiple users are randomly deployed in the indoor environments.}
\label{system_model}
\vspace{-0.1in}
\end{figure}

As depicted in Fig.~\ref{system_model}, we consider an UL PASS consisting of multiple randomly deployed users, each equipped with a single antenna. In this article, the dimensions of the room are defined as $x \in \left[ {0,D} \right], y \in \left[ {0,D} \right], z \in \left[ {0,h} \right]$. An AP is positioned at the corner of the room's ceiling, while a single waveguide is installed along the edge of the ceiling, with its location defined as $W = \left[ {\left( {0,D} \right),0,h} \right]$. Note that although the volume and shape of the waveguide can generate standing waves that degrade its performance, for simplicity, this article assumes an ideal waveguide model.
The PAs are strategically placed on the waveguide to serve multiple randomly deployed users, with their locations optimally chosen to maximize the users' channel gains. It is assumed that the position of the $i$-th PA and the $i$-th user are denoted by ${L_i} = \left[ {{x_{li}},0,h} \right]$ and ${U_i} = \left[ {{x_{ui}},{y_{ui}},0} \right]$, respectively.

For the large-scale channel, to facilitate theoretical exploration of fundamental performance limits and asymptotic behavior, we begin by focusing our analysis on LoS channels, which are modelled as follows:
\begin{equation}
h({l_i},{u_i}) = \frac{c}{{4\pi {f_c}\left| {{U_i} - {L_i}} \right|}},
\end{equation}
where $c$ represents the speed of light, $f_c$ represents the career frequency, $\left| {{U_i} - {L_i}} \right| $ represents the distance between the $i$-th PA and the $i$-th user, which can be expressed as
\begin{equation}\label{distance}
\left| {{U_i} - {L_i}} \right| = \sqrt {{{\left( {{x_{li}} - {x_{ui}}} \right)}^2} + y_{ui}^2 + {h^2}} .
\end{equation}

In this article, we assume that multiple UL users are randomly distributed across the floor, with their positions uniformly distributed along the x-axis and y-axis. By adopting an one-dimensional stochastic geometry model, the coordinates $x$ and $y$ of each user can be expressed as follows:
\begin{equation}\label{stochastic geometry}
\begin{aligned}
& {f_x}\left( r \right) = {f_y}\left( r \right) \\
& = \frac{1}{{\sqrt {({D^2} + {h^2}) - {h^2}} }} = \frac{1}{D},{\rm{if}}\;\;h < r < \sqrt {({D^2} + {h^2})} .
\end{aligned}
\end{equation}

For the small-scale channel model, the channel gain is assumed to be normalized, considering only the phase shift induced by the propagation distance in both the waveguide and free-space environments, which can be simplified to:
\begin{equation}\label{small-scale channels}
\begin{aligned}
&\exp \left( { - \frac{{j2\pi \left| {{U_i} - {L_i}} \right|}}{\lambda }} \right)\exp \left( { - \frac{{j2\pi {x_i}}}{\lambda }} \right) \\
&= \exp \left( { - \frac{{j2\pi \left( {\left| {{U_i} - {L_i}} \right| + {x_i}} \right)}}{\lambda }} \right),
\end{aligned}
\end{equation}
where $\lambda$ represents the wave length.

\subsection{Signal Model}

In the UL transmission, we consider the most representative technique for allocating wireless resources to users, i.e., time division multiple access. Thus, the total available time resource block is divided into $I$ time slots, with each time slot exclusively assigned to one user. Thus, the received signal by the PA can then propagate through the waveguide, and the received signal of the $i$-th user at the AP can be expressed as:
\begin{equation}\label{received signal at AP}
{y_i} = h({l_i},{u_i})\exp \left( { - \frac{{j2\pi {\left( {\left| {{U_i} - {L_i}} \right| + {x_i}} \right)} }}{\lambda }} \right) \sqrt{ {P_i}}{s_i} + {\eta _i},
\end{equation}
where ${{P_i}}$ denotes the transmit power at user $i$, $\eta_i \sim \mathcal{N} (0, \sigma^2)$ is the additive white Gaussian noise (AWGN), and $ s_i $ represents the transmitted signal of the $i$-th user with $\mathbb{E}(s_i)=0$.

Note that in~\eqref{received signal at AP}, although the path loss within the waveguide is negligible due to its highly confined propagation environment, the phase shift induced during the signal transmission through the waveguide still needs to be accounted, which depends on the physical length of the waveguide and the frequency of the transmitted signal.

\section{Performance Analysis}

To thoroughly evaluate the performance benefits of the proposed PASS, we analyze three distinct scenarios:
\begin{itemize}
  \item i) Firstly, in the MPSU scenario, multiple PAs are deployed to serve their associated user, enabling enhanced signal quality. However, optimizing the positions of the PAs is essential to achieve higher antenna gains.
  \item ii) Secondly, recognizing the spatial constraints of deploying multiple PAs, a simpler SPSU scenario is considered, where a single PA is deployed to serve a single user. In this scenario, the PA is positioned near to its associated user to maximize performance.
    \item iii) Thirdly, when the number of PAs is limited, determining the optimal positions of PAs becomes crucial to maximize the ergodic sum rate of PASS.
\end{itemize}
The comparative analysis in this section provides a comprehensive understanding of how the deployment strategy impacts the network's overall efficiency and user performance.

\subsection{Using Multiple PAs for a Single User (MPSU)}
In this subsection, multiple PAs are deployed along the waveguide to serve multiple users, where $2N+1$ PAs are associated with each individual user, namely MPSU. For simplicity, the indices of the PAs associated with the $i$-th user are denoted as ${L_{i,{n}}}$ with $ n =  - N,...,0,...,N$. When $n=0$, the associated PA is positioned at ${L_{i,0}} = \left[ {{x_{ui}},0,h} \right]$. The PAs with $n<0$ are located closer to the AP and are referred as the near array (NA), while the PAs with $n>0$ are located farther from the AP, forming the far array (FA). The spacing between the $n$-th and the $n+1$ the PA is defined as $\Delta {x_n} = {x_{li,n + 1}} - {x_{li,n}}$.

The channel gain between the $n$-th PA and its associated user in the MPSU scenario can be rewritten as:
\begin{equation}\label{Massive PAs_channel}
\begin{aligned}
& h({l_{i,n}},{u_i}) \\
&  = \frac{c}{{4\pi {f_c}\left| {{U_i} - {L_{i,n}}} \right|}}\exp \left( { - \frac{{j2\pi \left( {\left| {{U_i} - {L_{i,n}}} \right| + {x_{li,n}}} \right)}}{\lambda }} \right).
\end{aligned}
\end{equation}

The channel vector between the PAs and the associated user is given by
\begin{equation}\label{massive PAs channel vector}
{\rm \bf{h}} = \left[ {h({l_{i, - n}},{u_i}); \ldots ;h({l_{i,n}},{u_i})} \right].
\end{equation}

Since the PAs are distributed on the waveguide, which can be treated as a linear array, the effective channel gain can be reformulated as:
\begin{equation}\label{effective channel gain massive PAs}
\begin{aligned}
&{\left| {\sum\limits_{ - n}^n {} h({l_{i,n}},{u_i})} \right|^2} \\
& = \gamma {\left| {\sum\limits_{ - n}^n {\frac{1}{{\left| {{U_i} - {L_{i,n}}} \right|}}\exp \left( { - \frac{{j2\pi {\left( {\left| {{U_i} - {L_{i,n}}} \right| + {x_{li,n}}} \right)}}}{\lambda }} \right)} } \right|^2}.
\end{aligned}
\end{equation}

When the locations of PAs are randomly distributed, deriving a closed-form solution for the summation in~\eqref{effective channel gain massive PAs} becomes infeasible. To address this challenge, we first determine an optimized solution where the signals received at each PA are coherently combined at the AP, ensuring maximum effective channel gain. The goal of optimization ensures that the phases of the signals arriving at the AP are aligned, thereby enhancing the effective channel gain, which can be written as:
\begin{equation}\label{cophase}
\begin{aligned}
&\arg \left( {\exp \left( { - \frac{{j2\pi \left( {\left| {{U_i} - {L_{i,n + 1}}} \right| + {x_{li,n + 1}}} \right)}}{\lambda }} \right)} \right) \\
& - \arg \left( {\exp \left( { - \frac{{j2\pi \left( {\left| {{U_i} - {L_{i,n}}} \right| + {x_{li,n}}} \right)}}{\lambda }} \right)} \right) = 2\pi .
\end{aligned}
\end{equation}

The optimized positions of PAs, which satisfy the constraint in~\eqref{cophase}, are derived in the following lemmas in the NZ and FZ schemes.

\begin{lemma}\label{lemma_1_new_Optimized massive PAs location}
In the UL transmission, when multiple PAs are deployed for an individual user in the LoS channels, if ${d_0} >  N \lambda$ holds, the optimized position of the $n$-th PA of the $i$-th user can be expressed as:
\begin{equation}\label{lemma_equ_massive_PAs_position_FA}
{x_n} = \frac{{n\lambda \left( {2{d_0} + n\lambda } \right)}}{{2\left( {{d_0} + n\lambda } \right)}}, n =  - N,...,0,...,N ,
\end{equation}
where ${d_0} = \sqrt {y_{ui}^2 + {h^2}} $.
\begin{proof}
  Please refer to Appendix A.
\end{proof}
\end{lemma}

We can observe from~\eqref{lemma_equ_massive_PAs_position_FA} that the position of each PA is a function of the PA index. Consequently, it becomes essential to analyze the spatial distribution of PAs, which is formally addressed in the following propositions for providing deeper insights into the placement and spacing of PAs, also ensuring optimal signal coherence and performance in the system. To begin with, we define the FZ case as the case where the reference distance ${d_0}$ is significantly greater than $N \lambda$, ensuring that the user is located far from the array. Conversely, if ${d_0}$ does not satisfy the above condition, the user is considered to be in the NZ.

\begin{proposition}\label{Propo_new1_non-uniform-distribution}
Based on the optimized position of multiple PAs in~\eqref{lemma_equ_massive_PAs_position_FA}, it is observed that the PAs obey a non-uniform distribution when user is located in the NZ scheme.
\begin{proof}
The proof starts by the derivation of~\eqref{lemma_equ_massive_PAs_position_FA}, which can be given by:
\begin{equation}\label{propro new1 equa derivative}
\begin{aligned}
\frac{{d{x_n}}}{{dn}} &= \frac{{\left( {2\lambda {{\left( {{d_0} + n\lambda } \right)}^2}} \right) + 2\lambda d_0^2}}{{{{\left( {2\left( {{d_0} + n\lambda } \right)} \right)}^2}}}
= \frac{\lambda }{2} + \frac{{\lambda d_0^2}}{{2{{\left( {{d_0} + n\lambda } \right)}^2}}}.
\end{aligned}
\end{equation}
It is evident that \eqref{propro new1 equa derivative} is a monotonically decreasing function with respect to $n$, indicating that the spacing between PAs is not constant when $n$ is not large enough. As a result, the distance distribution of PAs is non-uniform, and this completes the proof.
\end{proof}
\end{proposition}

\begin{proposition}\label{Propo_new2_asymmetric-distribution}
Based on the optimized position of multiple PAs derived in~\eqref{lemma_equ_massive_PAs_position_FA}, it is well established that the spatial distribution of PAs in the NA and FA is asymmetric in the NZ scheme, which arises due to the non-uniform distance distribution of PAs.
\begin{proof}
When $\left| n \right|\lambda {d_0} \ne 0$, it is obvious that $\frac{{\left| n \right|\lambda \left( {2{d_0} - \left| n \right|\lambda } \right)}}{{2\left( {{d_0} - \left| n \right|\lambda } \right)}} \ne \frac{{\left| n \right|\lambda \left( {2{d_0} + \left| n \right|\lambda } \right)}}{{2\left( {{d_0} + \left| n \right|\lambda } \right)}}$ holds, which indicates that the NA and the FA are asymmetric distributed, thus the proof is complete.
\end{proof}
\end{proposition}

\begin{figure}[t!]
\centering
\includegraphics[width =2.5 in]{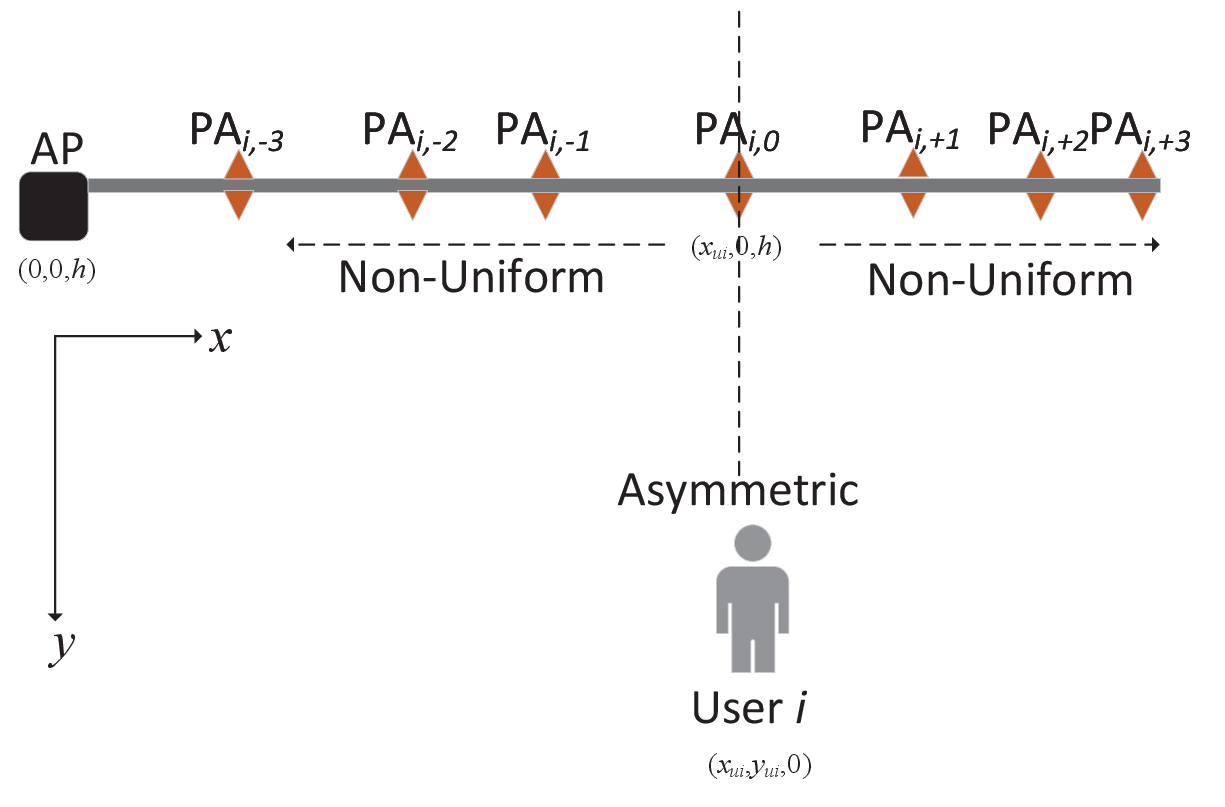}
\caption{Top view of the asymmetric non-uniform PAs in the MPSU scenario.}
\label{massive_PAs_illu_fig}
\vspace{-0.1in}
\end{figure}

\begin{remark}\label{remark_new1_non-symmetry_non_uniform_dis}
Based on the optimized positions of multiple PAs derived in~\eqref{lemma_equ_massive_PAs_position_FA}, when the distance between the PAs and the user is not sufficiently large, and supported by the results in {\textbf{Proposition~\ref{Propo_new1_non-uniform-distribution}} and \textbf{Proposition~\ref{Propo_new2_asymmetric-distribution}}}, the optimized locations of PAs exhibit an {\textbf{asymmetric non-uniform distribution}}. For a clearer understanding, the asymmetric non-uniform distribution of PAs is visually represented in Fig.~\ref{massive_PAs_illu_fig}, highlighting the spatial characteristics of the optimized PA positions in the NZ case.
\end{remark}

We then turn our attention to the FZ case in the following proposition, where ${d_0} \gg N\lambda $ holds.

\begin{proposition}\label{Propo_new3_far-field approach}
Based on the optimized position of multiple PAs in~\eqref{lemma_equ_massive_PAs_position_FA}, when the distance between the PAs and user is large enough, i.e. ${d_0} \gg N\lambda $, the position of the $n$-th PA in the FZ scheme can be approximated to
\begin{equation}\label{propo_new3_equa}
{x_n} = n\lambda .
\end{equation}
\begin{proof}
When ${d_0} \gg N\lambda $ holds, we have ${x_n} \approx \frac{{n\lambda 2{d_0}}}{{2{d_0}}} = n\lambda $, and the proof is complete.
\end{proof}
\end{proposition}

\begin{remark}\label{remark_new2_far-field-symmetric uniform}
When the distance between the PAs and user is large enough, and based on the approximated position of multiple PAs in {\textbf{Proposition~\ref{Propo_new3_far-field approach}}}, the approximated location of PAs follows a {\textbf{symmetric uniform distribution}} in the FZ scheme.
\end{remark}

By utilizing the above well-designed position of multiple PAs, the effective channel gain in~\eqref{effective channel gain massive PAs} can be reformulated as:
\begin{equation}\label{effective channel gain opt expre}
\begin{aligned}
{\left| {\sum\limits_{n =  - N}^N {} h({l_{i,n}},{u_i})} \right|^2} &= \gamma {\left| {\sum\limits_{n =  - N}^N {\frac{1}{{\left| {{U_i} - {L_{i,n}}} \right|}}} } \right|^2}\\
& = \gamma {\left| {\sum\limits_{n =  - N}^N {\frac{1}{{\sqrt {{d_0} + {x_n}} }}} } \right|^2}.
\end{aligned}
\end{equation}

It is obvious that the summation of~\eqref{effective channel gain opt expre} is not solvable. Therefore, the summation is approximated by transforming it into an integral, which can be expressed as follows:
\begin{equation}\label{effective channel integral_oring}
\sum\limits_{n =  - N}^N {} h({l_{i,n}},{u_i}) = \int_{ - N}^N {\frac{1}{{\sqrt {d_0^2 + {{\left( {\frac{{n\lambda \left( {2{d_0} + n\lambda } \right)}}{{2\left( {{d_0} + n\lambda } \right)}}} \right)}^2}} }}dn}
\end{equation}

Again, it is evident that the closed-form expression of~\eqref{effective channel integral_oring} is not derivable in the NZ scheme. Therefore, we only consider the FZ scheme in the following Lemma, where ${\left( {\frac{{n\lambda \left( {2{d_0} + n\lambda } \right)}}{{2\left( {{d_0} + n\lambda } \right)}}} \right)^2} \approx {n^2}{\lambda ^2}$.

\begin{lemma}\label{lemma_2_new_effective channel gain in far zone}
In the UL transmission, when multiple PAs are deployed for an individual user in the LoS channels, if ${d_0} \gg n\lambda $ holds, the effective channel gain of the $i$-th user in the FZ scheme can be expressed as:
\begin{equation}\label{lemma_equ_massive_PAs_channel gain closed-form}
\left| {\sum\limits_{n =  - N}^N {} h({l_{i,n}},{u_i})} \right| = \frac{2}{\lambda }\ln \left( {\frac{{N\lambda }}{{{d_0}}} + \sqrt {1 + {{\left( {\frac{{N\lambda }}{{{d_0}}}} \right)}^2}} } \right).
\end{equation}
\begin{proof}
We first rewrite the integral in~\eqref{effective channel integral_oring} as follows:
\begin{equation}\label{Appen NewB integral rewrite}
\int_{ - N}^N {\frac{1}{{\sqrt {d_0^2 + {n^2}{\lambda ^2}} }}dn}  = 2\int_0^N {\frac{1}{{\sqrt {d_0^2 + {n^2}{\lambda ^2}} }}dn} .
\end{equation}

Then, the closed-form expression in the FZ scheme can be obtained as follows:
\begin{equation}\label{Appen newB_integral result}
\left| {\sum\limits_{n =  - N}^N {} h({l_{i,n}},{u_i})} \right| = \frac{2}{\lambda }\ln \left( {\frac{{N\lambda d_0^{} + d_0^{}\sqrt {d_0^2 + {N^2}{\lambda ^2}} }}{{d_0^{}}}} \right).
\end{equation}
By some algebraic manipulations, the results in~\eqref{lemma_equ_massive_PAs_channel gain closed-form} can be obtained, and the proof is complete.
\end{proof}
\end{lemma}

By utilized the above optimized position of multiple PAs, the signals can be coherent at the AP, which can be treated as an ideal maximal ratio combination. Thus, the signal-to-noise ratio (SNR) of the $i$-th user can be formulated by the derived channel gain in~\eqref{lemma_equ_massive_PAs_channel gain closed-form} as follows:
\begin{equation}\label{SNR_massive PAs}
\begin{aligned}
&\frac{{{{\left| {\sum\limits_{n =  - N}^N {} h({l_{i,n}},{u_i})} \right|}^2}{P_i}}}{{\sum\limits_{n =  - N}^N {} \left| {{\sigma ^2}} \right|}} \\
& \approx \frac{{2\gamma {{P_i}}  {{\left( {\ln \left( {\frac{{N\lambda }}{{d_0^{}}} + \sqrt {1 + {{\left( {\frac{{N\lambda }}{{d_0^{}}}} \right)}^2}} } \right)} \right)}^2}}}{{N{\lambda ^2}}},
\end{aligned}
\end{equation}
where $\sigma^2 $ represents the power of AWGN. For notation simplicity, we set $\gamma  \buildrel \Delta \over = {\left( {\frac{c}{{4\pi {f_c}\sigma }}} \right)^2}$.

When the users are random deployed in UL transmission, the closed-form expression of the term ${\log _2}\left( {1 + \frac{{2\gamma {{\left( {\ln \left( {\frac{{N\lambda }}{{d_0^{}}} + \sqrt {1 + {{\left( {\frac{{N\lambda }}{{d_0^{}}}} \right)}^2}} } \right)} \right)}^2}}}{{N{\lambda ^2}}}} \right)$ is non-solvable due to the fact that the inverse function does not exist. Therefore, we again consider a FZ approximation, where $d_0^{} \gg N\lambda $.

In this scenario, the $0$-th PA is strategically placed near to the $i$-th user to minimize path loss. Consequently, the horizontal distance between the PA and the user is zero, i.e., ${x_{li,0}} - {x_{ui}} = 0$. Thus, the effective distance between the $0$-th PA and its associated user is simplified to

\begin{equation}\label{Multiple PAs OMA}
\left| {{U_i} - {L_{i,0}}} \right| = \sqrt {y_{ui}^2 + {h^2}} .
\end{equation}

\begin{theorem}\label{theorem New1 massive PAs ergodic rate}
In the UL transmission, when multiple PAs are deployed for an individual user in the LoS channels, the ergodic rate of the $i$-th user in the MPSU scenario can be expressed as:
\begin{equation}\label{theorem new1_equ_massive_PAs}
\begin{aligned}
&R_{i,{\rm{mPA}}s} = \frac{1}{I}{\log _2}\left( {1 + \frac{{2N\gamma {P_i}}}{{D_{}^2 + {h^2}}}} \right) + \frac{2}{{ID\ln 2}}\\
&\times \left( {\sqrt {{h^2} + 2N\gamma {P_i}} \arctan \frac{D}{{\sqrt {{h^2} + 2N\gamma {P_i}} }} - h\arctan \frac{D}{h}} \right).
\end{aligned}
\end{equation}
\begin{proof}
Please refer to Appendix B for more details.
\end{proof}
\end{theorem}

We now turn our attention to the high-SNR regime approach, where the transmit SNR is sufficiently large, i.e., $\frac{{{P_i}}}{{{\sigma ^2}}} \to \infty $, resulting in $\arctan x \to x$. Therefore, the high-SNR approximation is derived in the following proposition.
\begin{corollary}\label{corro2_new_MUSU_high-SNR_appro}
In the UL transmission, when multiple PAs are deployed for each user in the MPSU scenario, and the transmit SNR is high enough, the approximate achievable ergodic rate in LoS channels of the $i$-th user in the MPSU scenario can be given by:
\begin{equation}\label{coro2_new_equ}
\begin{aligned}
&\bar R_{i,{\rm{mPA}}s} \to \frac{1}{I}{\log _2}\left( {1 + \frac{{2N\gamma {P_i}}}{{D_{}^2 + {h^2}}}} \right){\rm{ }} \\
&+ \frac{2}{{I\ln 2}} - \frac{{2h}}{{ID\ln 2}}\arctan \frac{D}{h}.
 \end{aligned}
\end{equation}
\begin{proof}
By substituting $\mathop {\lim }\limits_{x \to 0} {\rm{ }}\arctan x \to x$ in the high-SNR regime, the results in~\eqref{coro2_new_equ} can be readily obtained.
\end{proof}
\end{corollary}

\begin{remark}\label{remark_new3_MUSU gain}
Based on the results in~\eqref{theorem new1_equ_massive_PAs} and its high-SNR approximation in~\eqref{coro2_new_equ}, it can be observed that the antenna gain in the MPSU scenario asymptotically approaches $2N$ in high-SNR regime, which indicates that the deployment of PAs significantly enhances the channel gain, highlighting the potential of massive PAs deployments.
\end{remark}

To gain deep insights into the system’s performance, the high-SNR slope is worth estimating, which is considered as the key parameter determining the ergodic rate in the high-SNR regime. To do so, we first define the high-SNR slope as
\begin{equation}\label{high-SNR slope defination}
{\Lambda _{i,{\rm{mPA}}s}} = \mathop {\lim }\limits_{\frac{{{P_i}}}{{{\sigma _2}}} \to \infty } \frac{{{R_{i,{\rm{mPA}}s}}}}{{{{\log }_2}\left( {1 + \frac{{{P_i}}}{{{\sigma _2}}}} \right)}}.
\end{equation}

\begin{proposition}\label{high-SNR slope MPSU}
In the UL transmission, when multiple PAs are deployed for each user in the MPSU scenario, and the transmit SNR is high enough, the high-SNR slope of the $i$-th user in the MPSU scenario can be given by:
\begin{equation}\label{high-SNR MPSU}
{\Lambda _{i,{\rm{mPA}}s}} = \frac{1}{I}.
\end{equation}
\begin{proof}
By utilizing L'Hospital's rule, it is obvious that $\mathop {\lim }\limits_{\frac{{{P_i}}}{{{\sigma _2}}} \to \infty } \frac{{{{\log }_2}\left( {1 + \frac{{2N\gamma {P_i}}}{{D_{}^2 + {h^2}}}} \right)}}{{{{\log }_2}\left( {1 + \frac{{{P_i}}}{{{\sigma _2}}}} \right)}} \to 1$ and $\mathop {\lim }\limits_{\frac{{{P_i}}}{{{\sigma _2}}} \to \infty } \frac{{2\sqrt {{h^2} + 2N\gamma {P_i}} }}{{ID\ln 2}}\arctan \frac{D}{{\sqrt {{h^2} + 2N\gamma {P_i}} }} \to 0$.

Thus, the high-SNR slope can be formulated as
\begin{equation}\label{high-SNR-SLOPE final}
{\Lambda _{i,{\rm{mPA}}s}} = \frac{1}{I}+0=\frac{1}{I}.
\end{equation}
The proof is complete.
\end{proof}
\end{proposition}

\begin{remark}\label{remark_new4_MUSU hihg-SNR slope}
Based on the results in~\eqref{high-SNR MPSU}, it can be observed that the high-SNR slope in the MPSU scenario approaches $\frac{1}{I}$. This phenomenon suggests that the high-SNR slope is primarily determined by the number of users, which remains unaffected by the number of PAs.
\end{remark}

\subsection{Using A Single PA for A Single User (SPSU)}

It is worth noting that in wireless communications, the wavelength is typically at the centimeter level, making the deployment of massive PAs for each user in indoor environments impractical. Therefore, we then pay our attention on the second scenario, in which a single PA is deployed for each users, namely SPSU, which is a special scenario of MPSU. Consequently, our focus shifts to the second SPSU scenario, where a single PA is deployed for each user, represents a simplified and practical subset of the MPSU scenario. By examining the SPSU scenario, we aim to explore its potential and performance under more feasible deployment conditions.

To enable access services for multiple users, each PA is activated sequentially within its allocated time slot. Consequently, the received SNR for the $i$-th user at the AP can be expressed as:
\begin{equation}\label{UL_OMA_MulPAs_SNR}
SN{R_i} = \frac{{{{\left| {h({l_i},{u_i})} \right|}^2}{P_i}}}{{{\sigma ^2}}} = \frac{{\gamma {P_i}}}{{y_{ui}^2 + {h^2}}}.
\end{equation}

The data rate of the $i$-th user in SPSU scenario can be rewritten as
\begin{equation}\label{OMA_MulPAs_rate}
\begin{aligned}
R_{i,{\rm{PA}}s} &= \frac{1}{I}{\log _2}\left( {1 + \frac{{\gamma {P_i}}}{{y_{ui}^2 + {h^2}}}} \right) \\
& = \frac{1}{I}{\log _2}\left( {y_{ui}^2 + {h^2} + \gamma {P_i}} \right) - \frac{1}{I}{\log _2}\left( {y_{ui}^2 + {h^2}} \right).
\end{aligned}
\end{equation}

Then, the ergodic rate performance can be evaluated by considering the random process of users in the following Theorem.
\begin{theorem}\label{theorem1: OMA_Mul_PA}
In the UL transmission, when a single PA is deployed for each user in the LoS channels, the achievable ergodic rate of the $i$-th user in SPSU scenario can be expressed as:
\begin{equation}\label{Theo1_equ_OMA_UL_Mul_PA}
\begin{aligned}
& R_{i,{\rm{PA}}s}  = \frac{1}{I}{\log _2}\left( {1 + \frac{{\gamma {P_i}}}{{D_{}^2 + {h^2}}}} \right) + \\
&  \frac{2}{{ID\ln 2}}\left( {\sqrt {{h^2} + \gamma {P_i}} \arctan \frac{D}{{\sqrt {{h^2} + \gamma {P_i}} }} - h\arctan \frac{D}{h}} \right).
\end{aligned}
\end{equation}
\begin{proof}
Similar to Appendix B, the results in~\eqref{Theo1_equ_OMA_UL_Mul_PA} can be readily obtained.
\end{proof}
\end{theorem}

It is however quite challenging to directly obtain engineering insights from~\eqref{Theo1_equ_OMA_UL_Mul_PA} due to the inverse trigonometric function. Thus, in order to gain further insights, the approximate achievable ergodic rate is derived in the following Corollaries.

\begin{corollary}\label{Corro1_OMA_Mul_PA_approx}
Given the assumption that the height of the waveguide is greater than the size of the room, i.e., $h>D$. In the UL transmission, when a single PA is deployed for each user, the approximate achievable ergodic rate in LoS channels of the $i$-th user in SPSU scenario can be calculated as:
\begin{equation}\label{corro1 equati}
\begin{aligned}
& {\hat R}_{i,{\rm{PA}}s} = \frac{1}{I}{\log _2}\left( {1 + \frac{{\gamma {P_i}}}{{D_{}^2 + {h^2}}}} \right) \\
&+ \frac{2}{{I \ln 2}}\sum\limits_{k = 0}^\infty  {\frac{{{{\left( { - 1} \right)}^k}{D^{2k}}}}{{2k + 1}}} \left( {\frac{1}{{{{\left( {{h^2} + \gamma {P_i}} \right)}^k}}} - \frac{1}{{{h^{2k}}}}} \right).
\end{aligned}
\end{equation}
\begin{proof}
Please refer to Appendix C.
\end{proof}
\end{corollary}
It is important to noted that Corollary~\ref{Corro1_OMA_Mul_PA_approx} is applicable only in specific scenarios, such as stadium or auditorium, where the spatial dimensions and deployment configurations align with the assumptions. 

\begin{corollary}\label{corro2}
In the UL transmission, when multiple PAs are deployed for each user, and the transmit SNR is high enough, the approximate achievable ergodic rate in LoS channels of the $i$-th user can be given by:
\begin{equation}\label{coro2_equ}
\begin{aligned}
{\bar R}_{i,{\rm{PA}}s} \to \frac{1}{I}{\log _2}\left( {1 + \frac{{\gamma {P_i}}}{{D_{}^2 + {h^2}}}} \right) \\
 + \frac{2}{{I\ln 2}} - \frac{{2h}}{{ID\ln 2}}\arctan \frac{D}{h}.
 \end{aligned}
\end{equation}
\begin{proof}
By substituting $\mathop {\lim }\limits_{x \to 0} {\rm{ }}\arctan x \to x$ in the high-SNR regime, the results in~\eqref{coro2_equ} can be readily obtained.
\end{proof}
\end{corollary}

\begin{remark}\label{remark sec1_MuPA OMA}
Based on the results in~\eqref{coro2_equ}, and the fact $R_{{\rm{PA}}s} = \sum\limits_{i = 1}^I {} R_{i,{\rm{PA}}s}$, it becomes evident that when multiple PAs are deployed to serve their corresponding randomly located users, the ergodic sum rate of the UL transmission remains independent of the number of users.
\end{remark}

To demonstrate the advantages of MPSU scenario, we compare the ergodic sum rate performance between MPSU and SPSU scenarios in the following proposition.
\begin{proposition}\label{propo4new_MPSU_SPSU compare}
In the UL transmission, the ergodic sum rate of the MPSU scenario is strictly greater than that of the SPSU scenario.
\begin{proof}
In order to compare the considered MPSU and SPSU scenarios, we have
\begin{equation}\label{propo_new4_first compare}
\begin{aligned}
&R_{{\rm{mPA}}s} - R_{{\rm{PA}}s} = \sum\limits_{i = 1}^I {} R_{i,{\rm{mPA}}s} - \sum\limits_{i = 1}^I {} R_{i,{\rm{PA}}s} \\
& ={\log _2}\left( {1 + \frac{{2N\gamma {P_i}}}{{D_{}^2 + {h^2}}}} \right) - {\log _2}\left( {1 + \frac{{\gamma {P_i}}}{{D_{}^2 + {h^2}}}} \right) > 0.
\end{aligned}
\end{equation}
Since in the MPSU scenario, since the number of PAs is greater than 1, it is evident that the expression in~\eqref{propo_new4_first compare} is strictly greater than 0, which demonstrates the inherent performance gain of deploying multiple PAs, thereby completing the proof.
\end{proof}
\end{proposition}

\begin{proposition}\label{high-SNR slope SPSU}
In the UL transmission, when a single PA is deployed for each user in the SPSU scenario, the high-SNR slope of the $i$-th user in the SPSU scenario can be given by:
\begin{equation}\label{high-SNR SPSU}
{\Lambda _{i,{\rm{PA}}s}} = \frac{1}{I}.
\end{equation}
\begin{proof}
Similar to Proposition~\ref{high-SNR slope MPSU}, the results in~\eqref{high-SNR SPSU} can be readily obtained.
\end{proof}
\end{proposition}

\subsection{Using A Single PA for Multiple Users (SPMU)}

In this subsection, it is assumed that there is only a single PA deployed on the waveguide to serve multiple users. Unlike the previous MPSU and SPSU scenarios, where PAs are activated in separate time slots, a single PA is activated to serve all users across all time slots in SPMU scenario. The distance between PA and the $i$-th user is as defined in~\eqref{distance}.
In SPMU scenario, the PA is positioned at the center of the waveguide to optimize service to randomly deployed users, and the location of PA can therefore be expressed as:
\begin{equation}\label{location PA sing PA_OMA}
{L_{}} = \left[ {\frac{D}{2},0,h} \right].
\end{equation}

By doing so, the distance between PA and user $i$ can be rewritten as:
\begin{equation}\label{distance PA single_OMA}
\left| {{U_i} - L} \right| = \sqrt {{{\left( {\frac{D}{2} - {x_{ui}}} \right)}^2} + y_{ui}^2 + {h^2}} .
\end{equation}

Then, the achievable ergodic rate of the $i$-th user can be formulated as:
\begin{equation}\label{rate_singlePA_OMA}
\begin{aligned}
R_{i,{\rm{A1}}}& = \frac{1}{I}{\log _2}\left( {{{\left( {\frac{D}{2} - {x_{ui}}} \right)}^2} + y_{ui}^2 + {h^2}+ \gamma {P_i}} \right) \\
& - \frac{1}{I}{\log _2}\left( {{{\left( {\frac{D}{2} - {x_{ui}}} \right)}^2} + y_{ui}^2 + {h^2}} \right).
\end{aligned}
\end{equation}

\begin{theorem}\label{Theo2:single PA_OMA}
In the UL transmission, when a single PA is deployed on the center of waveguide in the LoS channels, the achievable ergodic rate of the $i$-th user in the SPMU scenario can be expressed as:
\begin{equation}\label{Theo2 equa_large_Inte}
\begin{aligned}
&\mathbb{ E} \left\{ {R_{i,{\rm{A1}}}} \right\} = \frac{1}{{{I D^2}}} \\
& \times \underbrace { \int_{ - \frac{D}{2}}^{\frac{D}{2}} {} \int_0^D {{{\log }_2}\left( {{{\left( {\frac{D}{2} - {x_{ui}}} \right)}^2} + y_{ui}^2 + {h^2} + \gamma {P_i}} \right)} dy_{ui}^{}d{x_{ui}}}_{{L_3}} \\
&- \underbrace {\frac{1}{{{I D^2}}}\int_{ - \frac{D}{2}}^{\frac{D}{2}} {} \int_0^D {{{\log }_2}\left( {{{\left( {\frac{D}{2} - {x_{ui}}} \right)}^2} + y_{ui}^2 + {h^2}} \right)} dy_{ui}^{}d{x_{ui}}}_{{L_4}}.
\end{aligned}
\end{equation}
\end{theorem}

Note that the exact analytical result of~\eqref{Theo2 equa_large_Inte} is highly complex and does not yield intuitive engineering insights. To address this issue, an approximation is derived in the following corollary.

\begin{corollary}\label{Corro3_OMA_Sing_PA_approx}
It is assumed that the height of waveguide is larger than the size of room, i.e., $h>D$. In the UL transmission, when only a single PA is deployed at the center of waveguide, the approximate achievable ergodic rate in LoS channels of the $i$-th user in SPMU can be given by:
\begin{equation}\label{corro3 equati}
\begin{aligned}
 {{\hat{ R}}_{i,{\rm{A1}}}}  & \approx \frac{1}{I}{\log _2}\left( {1 + \frac{{\gamma {P_i}}}{{2D_{}^2 + {h^2}}}} \right) \\
& - \frac{{\sqrt {4{I D^2} + {h^2}} }}{{D\ln 2}}\arctan \frac{D}{{\sqrt {4{D^2} + {h^2}} }}.
\end{aligned}
\end{equation}
\begin{proof}
Please refer to Appendix D.
\end{proof}
\end{corollary}

To demonstrate the advantages of employing multiple PAs, we compare the ergodic sum rate performance between the SPSU scenario and SPMU scenario.
\begin{proposition}\label{propo1_OMA compare}
In the UL transmission, when deploying multiple PAs on the waveguide, the ergodic sum rate of the SPSU scenario is strictly greater than that of the SPSU scenario.
\begin{proof}
Please refer to Appendix E.
\end{proof}
\end{proposition}

\begin{remark}\label{remark section II single PA OMA}
On the one hand, when a sufficient number of PAs are deployed, the considered SPSU scenario achieves a higher ergodic sum rate by minimizing the distance between each PA and its corresponding user.
On the other hand, the considered SPMU scenario offers a simpler implementation, particularly when the number of available PAs is limited.
\end{remark}

\begin{proposition}\label{high-SNR slope SPMU}
In the UL transmission, when a single PA is deployed at the center of waveguide for multiple users in the SPMU scenario, the high-SNR slope of the $i$-th user in the SPMU scenario can be given by:
\begin{equation}\label{high-SNR SPMU}
{\Lambda _{i,{\rm{A1}}}} = \frac{1}{I}.
\end{equation}
\begin{proof}
Similar to Proposition~\ref{high-SNR slope MPSU}, the results in~\eqref{high-SNR SPMU} can be readily obtained.
\end{proof}
\end{proposition}

\begin{remark}\label{remark high-SNR-slope-compare}
It is observed from \textbf{Remark~\ref{high-SNR slope MPSU},Remark\ref{high-SNR SPSU}, and Remark\ref{high-SNR SPMU}} that the high-SNR slopes of MPSU, SPSU, and SPMU are all~$\frac{1}{I}$, which indicates that the high-SNR slope is not impacted by the number of PAs or the number of users.
\end{remark}

\subsubsection{Optimized PA position for SPMU Scenario}

Based on the insights from the SPMU scenario, another question raises: when multiple users share a limited number of PAs, how can the optimal positions for the PAs be determined? For simplicity, we consider a two-user scenario, where the position of PA is defined by ${L_{}} = \left[ {x,0,h} \right]$. Then, the distance between PA and the two users are given by $\left| {{U_1} - L} \right| = \sqrt {{{\left( {x - {x_{u1}}} \right)}^2} + y_{u1}^2 + {h^2}} $ and $\left| {{U_2} - L} \right| = \sqrt {{{\left( {x - {x_{u2}}} \right)}^2} + y_{u2}^2 + {h^2}} $, respectively.
We then explore the closed-form expression of the optimized position in the following theorem.

\begin{theorem}\label{Theo2+1:single PA_OMA_opt}
In the UL transmission, when two available users are located on the floor and a single PA is deployed on the waveguide in the LoS channels, the optimized position ratio of PA is given by:
\begin{equation}\label{Theo2+1 equa_large_Inte_single PA opt}
\begin{aligned}
\frac{{\left( {x - {x_{u1}}} \right)}}{{\left( {{x_{u2}} - x} \right)}} = \frac{{{{\left( {x - {x_{u1}}} \right)}^2} + y_{u1}^2 + {h^2}}}{{{{\left( {x - {x_{u2}}} \right)}^2} + y_{u2}^2 + {h^2}}}.
\end{aligned}
\end{equation}
\begin{proof}
Please refer to Appendix F for more details.
\end{proof}
\end{theorem}

\begin{corollary}\label{Corro3+1OMA_Sing_PA_position opt}
In the UL transmission, when two available users hold $y_{u1}=y_{u2}$, and a single PA is deployed on the waveguide in the LoS channels, the optimized position of PA is given by:
\begin{equation}\label{corro3+1 equati_OMA_Sing_PA_position opt}
\begin{aligned}
x = \frac{{{x_{u1}} + {x_{u2}}}}{2}.
\end{aligned}
\end{equation}
\begin{proof}
By substituting $y_{u1}=y_{u2}$, the results in~\eqref{corro3+1 equati_OMA_Sing_PA_position opt} can be readily obtained.
\end{proof}
\end{corollary}

\begin{remark}\label{remark section II single PA OMA_opt position}
Based on the results in~\eqref{Theo2+1 equa_large_Inte_single PA opt}, it can be observed that the optimal position of the PA is influenced by the distance ratio among multiple users, which highlights the importance of considering user spatial distribution when determining the optimal placement of PAs.
\end{remark}

\begin{remark}\label{remark section II single PA OMA_opt position2 equal y}
Based on the results in~\eqref{corro3+1 equati_OMA_Sing_PA_position opt}, it can be concluded that when two OMA users share the same $y$-axis coordinate, the optimal position of the PA is located at the midpoint of the $x$-axis coordinates of the two users.
\end{remark}

\section{Numerical Results}

In this section, numerical results are provided for the performance evaluation of the proposed PASS. Monte Carlo simulations are provided for verifying the accuracy of our analytical results. The carrier frequency is set to $f_c=2.4$ GHz. The transmission bandwidth of the proposed network is set to $BW=1$ MHz. In practice, the power of the AWGN is related to the bandwidth, which can be modeled as ${\sigma ^2} =  - 174 + 10{\rm{lo}}{{\rm{g}}_{10}}(BW)$ dBm. For simplicity, the number of users is set to $I=2$.

\subsection{MPSU Scenario}

\begin{figure}[t!]
\centering
\includegraphics[width =2.5in]{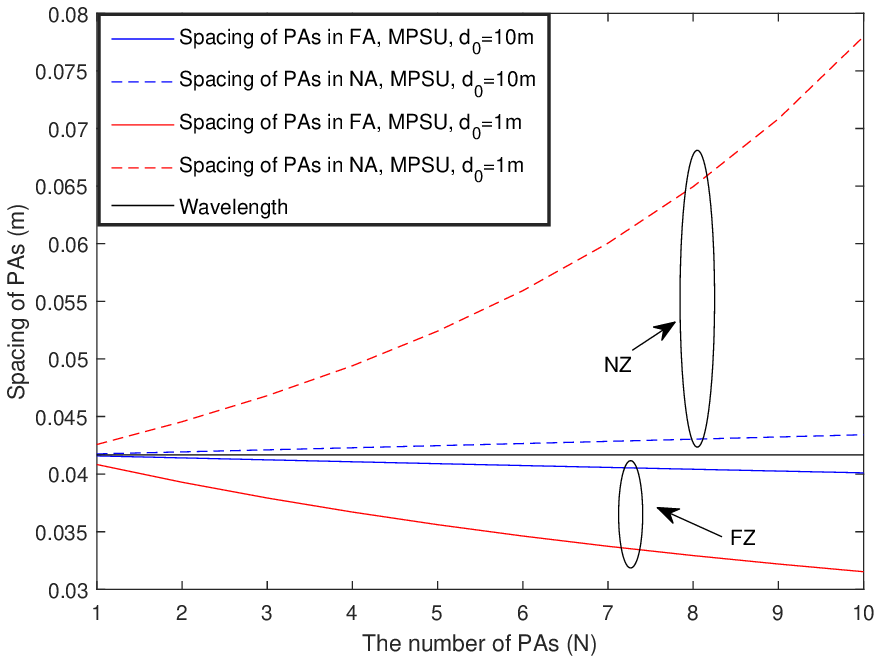}
\caption{Spacing of PAs in the MPSU scenario, where the analytical results are derived from~\eqref{lemma_equ_massive_PAs_position_FA}. The number of PAs is set to $N=10$.}
\label{Fig1 MPSU_PA spacing}
\vspace{-0.1in}
\end{figure}

\emph{1) Spacing of PAs in the MPSU scenario with different distances:} Fig.~\ref{Fig1 MPSU_PA spacing} illustrates the spacing of PAs of both NZ and FZ schemes in the MPSU scenario, where the number of PAs is set to $N=10$ for simplicity. The wavelength is also included as a benchmark scheme for comparison. In the NZ scheme, the spacing of PAs exhibits a pronounced difference between the NA and FA. Specifically, the spacing in the NA increases with the number of PAs, while the spacing in the FA decreases. This phenomenon indicates that the PAs are asymmetric non-uniform distribution in the NZ scheme. Conversely, in the FZ scheme, the spacing of PAs in the FA and NA remains approximately constant, with the FA spacing marginally exceeding the NA spacing, which is attributed to the relatively larger distance between the user and the PAs in the FZ scheme, leading to less significant variations in PA positioning. This phenomenon indicates that the PAs are distributed nearly to symmetric uniform distribution, which verifies our {\textbf{Remark~\ref{remark_new2_far-field-symmetric uniform}}}. Furthermore, as the number of PAs is high enough, the spacing of PAs of FA begins to converge toward the wavelength threshold. However, the PA spacing of NA increase, which also emphasizes the physical limitations imposed by the wavelength in designing PA spacing for MPSU scenarios.

\begin{figure}[t!]
\centering
\includegraphics[width =2.5in]{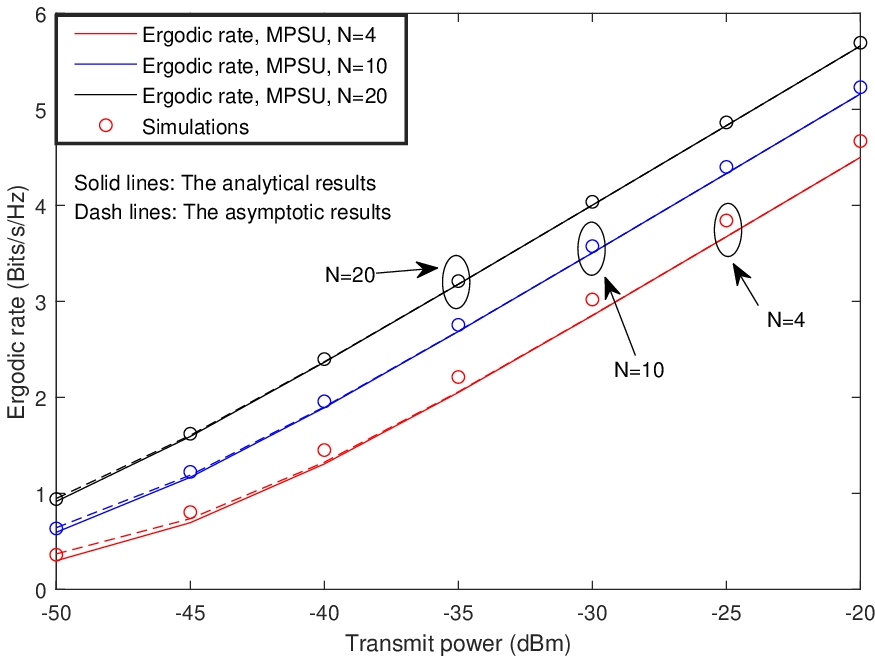}
\caption{Ergodic rate of the $i$-th user in the MPSU scenario, where the analytical results and asymptotic results are derived from~\eqref{theorem new1_equ_massive_PAs} and~\eqref{corro2_new_MUSU_high-SNR_appro}, respectively.}
\label{Fig1 MPSU_verify}
\vspace{-0.1in}
\end{figure}

\emph{2) Ergodic rate in the MPSU scenario with different number of PAs:} In Fig.~\ref{Fig1 MPSU_verify}, we evaluate the ergodic rate of the $i$-th user in the MPSU scenario, where the height and the dimension of the room are set to $h=20$ m and $D=10$ m, respectively. The number of PAs for each user is set to $N=4, 10, 20$. The solid curves and the dashed curves represent the analytical results and the asymptotic results, respectively. Simulation results are marked with circles for validation. As can be seen from black, blue and red curves, when the number of PAs is high enough, the gap between the analytical and asymptotic results diminishes, which verifies our analysis in {\textbf{Theorem~\ref{theorem New1 massive PAs ergodic rate}}} and {\textbf{Corollary~\ref{corro2_new_MUSU_high-SNR_appro}}}. This is because, in the analytical results, the signal between the user and its nearest PA is neglected when the number of PAs is sufficiently large. As a result, we can observe from the red curves that the gap between the analytical and simulation results becomes relatively significant when $N=4$, where the number of PAs is still comparatively low. It can be observed that as the number of PAs increases, the ergodic rate performance improves significantly, demonstrating the advantage of deploying multiple PAs in the MPSU scenario. This phenomenon corroborates the theoretical results that suggest an increased antenna gain with more PAs, which verified our {\textbf{Remark~\ref{remark_new3_MUSU gain}}}.

\begin{figure}[t!]
\centering
\includegraphics[width =2.5in]{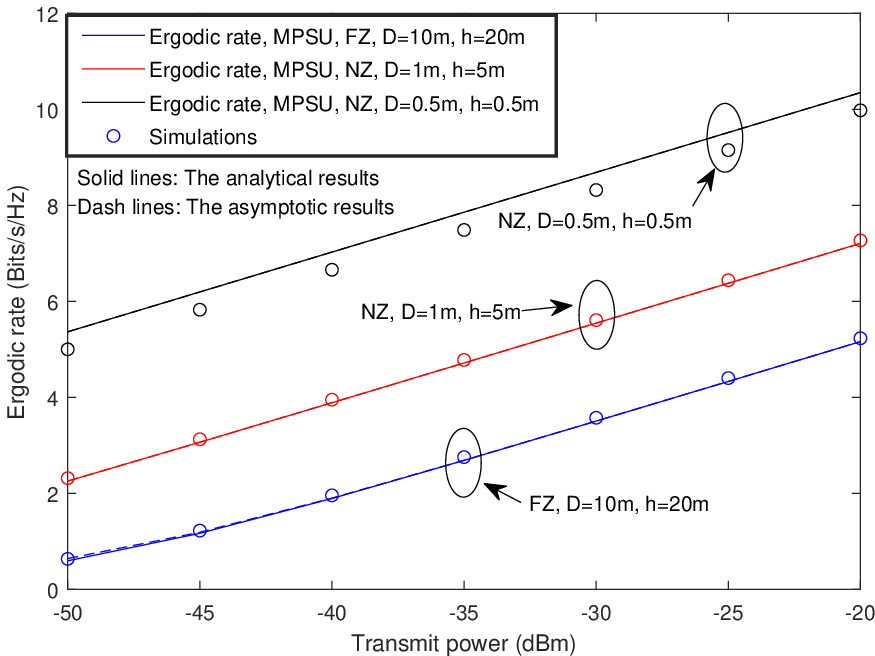}
\caption{Ergodic rate of the $i$-th user in the MPSU scenario in both NZ and FZ cases. The number of PAs is set to $N=10$.}
\label{Fig2 MPSU_impact_NZFZ}
\vspace{-0.1in}
\end{figure}

\emph{3) Ergodic rate in the MPSU scenario impacted by NZ and FZ cases:}
We then evaluate the impact of NZ and FZ cases in the MPSU scenario in Fig.~\ref{Fig2 MPSU_impact_NZFZ}, where the number of PAs is set to $N=10$. The NZ and FZ cases are differentiated by varying the waveguide heights and room dimensions. For comparison, the performance in the FZ case with $h=20$ m and $D=10$ m is provided as a benchmark scheme. In the NZ case, a small performance gap between the analytical and simulation results is observed, primarily due to the ergodic rate approximation being based on the FZ case. This gap narrows further in the high-SNR regime, with the ergodic rate difference approaching 0.2 bits/s/Hz, which becomes negligible as the SNR increases.
Moreover, we can also observe that when $h=5$ m and $D=1$ m, the distance between PAs and the user is much greater than $N\lambda$, validating that the FZ approximation is applicable in most practical scenarios. In addition, as can be observed from the red curves, when the minimum distance exceeds 5 meters, the analytical results are closely aligned with the simulation results in the NZ scheme, which indicates that the proposed analytical results can be deployed for the most application scenarios in the indoor environments.

\begin{figure}[t!]
\centering
\includegraphics[width =2.5in]{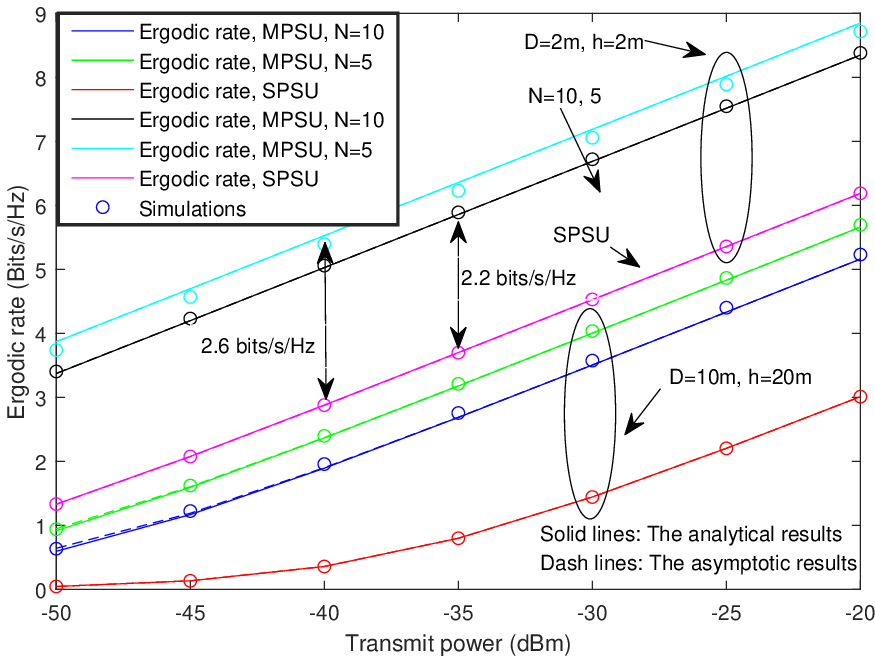}
\caption{Ergodic rate comparison between MPSU and SPSU scenarios.}
\label{Fig3 OMA MPSU_SPSU_compare}
\vspace{-0.1in}
\end{figure}

\emph{4) Ergodic Rate comparison between MPSU and SPSU scenarios:} Fig.~\ref{Fig3 OMA MPSU_SPSU_compare} illustrates the ergodic rate performance comparison between the MPSU and SPSU scenarios with different parameters. From the figure, it is evident that MPSU consistently outperforms SPSU, with the ergodic rate increasing as the number of PAs grows. Specifically, a notable performance gap of approximately 2.6 bits/s/Hz is observed between the $N=10$ and $N=5$ cases for MPSU. Furthermore, the ergodic rate gap between MPSU and SPSU at $N=5$ is approximately 2.2 bits/s/Hz, highlighting the significant advantage of deploying multiple PAs. Overall, it is clearly shown that the MPSU outperforms that of SPSU in terms of ergodic rate, which validates our {\textbf{Remark~\ref{propo4new_MPSU_SPSU compare}}} for the PASS. Additionally, the figure demonstrates the impact of spatial parameters on system performance, such as room dimensions and waveguide heights. When $D=2$ m and $h=2$ m, the MPSU scenario achieves a notably higher ergodic rate compared to the larger room dimension, which illustrates the importance of spatial optimization in maximizing the potential of PASS in indoor environments.

\subsection{SPSU Scenario}

\begin{figure}[t!]
\centering
\includegraphics[width =2.5in]{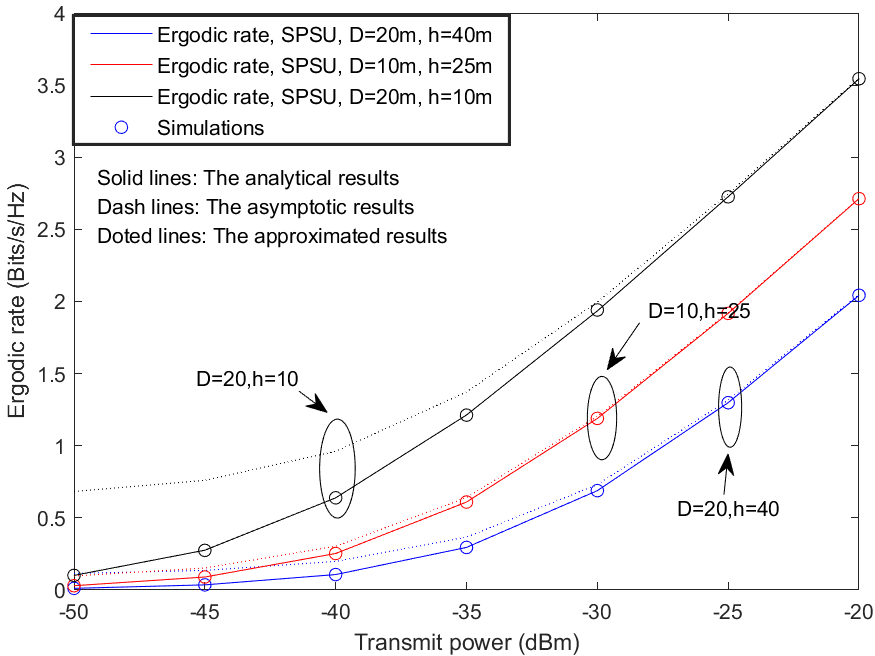}
\caption{Ergodic rate of the $i$-th user in the SPSU scenario, where the analytical results, asymptotic results and approximated results are derived from~\eqref{Theo1_equ_OMA_UL_Mul_PA},~\eqref{corro1 equati} and~\eqref{coro2_equ}, respectively.}
\label{Fig2 OMA Multi PAs}
\vspace{-0.1in}
\end{figure}

\emph{5) Ergodic Rate in the SPSU Scenario:} In Fig.~\ref{Fig2 OMA Multi PAs}, we evaluate the ergodic rate in the SPSU scenario, where a single PA is located on the waveguide to serve a single user. The closed agreement between analytical results, asymptotic results and simulation results verifies our correctness. It can be observed that as the transmit power increases, the ergodic rate improves consistently across all parameter setups. Notably, the black curves achieve the highest ergodic rate due to the smaller waveguide height, which minimizes the path loss between PAs and users. Note that as we can observe from the black curves, the asymptotic results are not available due to the fact that the constraint $h>D$ is not satisfied. Therefore, the Maclaurin’s series of inverse trigonometric function approach infinity. Additionally, we observe that the impact of waveguide height is more significant than that of room size, which is due to the fact that users are randomly distributed within the room, while the waveguide height remains constant. Overall, Fig.~\ref{Fig2 OMA Multi PAs} highlights the impact of spatial parameters on the ergodic rate performance of the SPSU scenario, validating the proposed analytical framework for a wide range of indoor application scenarios. The gained insights provide valuable guidance for optimizing PA placement and system design in practical deployments.

\begin{figure}[t!]
\centering
\includegraphics[width =2.5in]{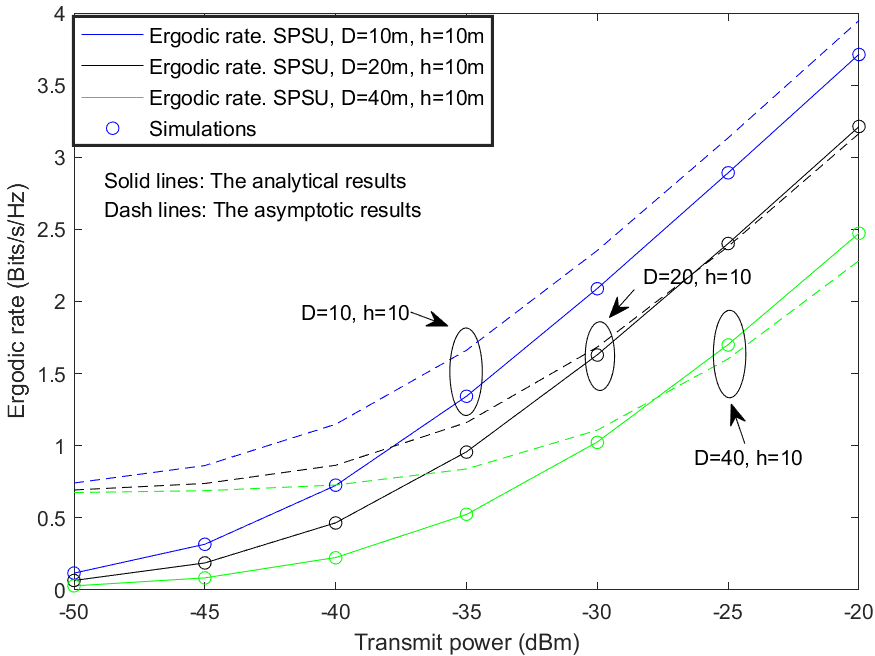}
\caption{Ergodic rate of the $i$-th user in the SPSU scenario, where the analytical results and asymptotic results are derived from~\eqref{Theo2 equa_large_Inte} and~\eqref{corro3 equati}, respectively.}
\label{fig3_OMA_Single PA}
\vspace{-0.1in}
\end{figure}

\emph{6) Ergodic rate in the SPSU scenario:} In Fig.~\ref{fig3_OMA_Single PA}, we evaluate the ergodic rate of the $i$-th user in UL transmission with a single PA versus different room dimension while maintaining a constant waveguide height. The solid curves denote the analytical results, while the dashed curves correspond to the asymptotic results. Circular markers represent simulation results, which align closely with the theoretical findings. We can see that the gap between asymptotic results and analytical results is smaller than 0.2 bits/s/Hz, which affirms the accuracy of the derived asymptotic expressions. Note that the performance error between the asymptotic results and analytical results mainly depends on the approximation of~\eqref{AppenC_L42}.
Observe that the blue curve achieves the highest ergodic rate due to the smaller room dimension. Conversely, the green curve shows the lowest performance, attributed to the larger room dimension, which increases the path loss between users and PA.

\begin{figure}[t!]
\centering
\includegraphics[width =2.5in]{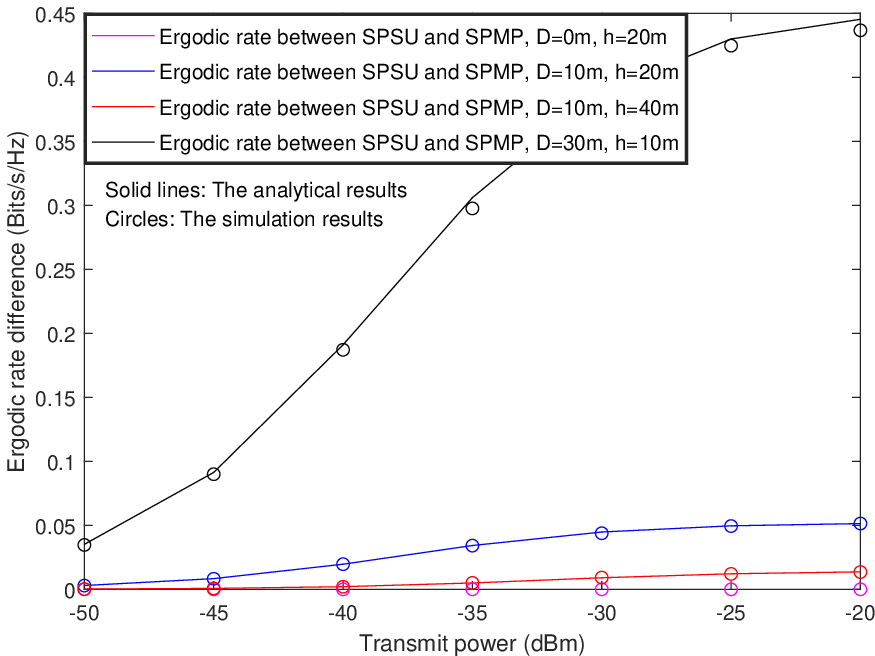}
\caption{Ergodic rate difference between SPSU and SPMU scenarios.}
\label{fig4_OMA_compare}
\vspace{-0.1in}
\end{figure}

\emph{7) Ergodic rate difference in SPSU and SPMU Scenarios:} In Fig.~\ref{fig4_OMA_compare}, we evaluate the ergodic rate difference between multiple-PAs scenario and single-PA scenario in UL transmission under varying room dimensions and waveguide heights. The analytical results is derived by~\eqref{Theo1_equ_OMA_UL_Mul_PA}-\eqref{Theo2 equa_large_Inte}, and the simulation results are obtained by the Mont Carlo simulations.
It is evident that the ergodic sum rate in the multiple-PAs scenario consistently outperforms than that of the single-PA scenario, corroborating our {\textbf{Proposition~\ref{propo1_OMA compare}}}, {\textbf{Remark~\ref{remark sec1_MuPA OMA}}} and {\textbf{Remark~\ref{remark section II single PA OMA}}}. This is because that in the multiple-PAs scenario, the distance between each PA and its corresponding user is minimized, which improves the received signal power. However, due to the deployment of additional PAs, the performance improvement comes at the cost of increased hardware complexity. Furthermore, all the curves highlight that the room dimensions and waveguide height significantly influence the ergodic rate difference among users in UL transmission, with larger rooms and lower waveguide leading to more pronounced performance gap.

\begin{figure}[t!]
\centering
\includegraphics[width =2.5in]{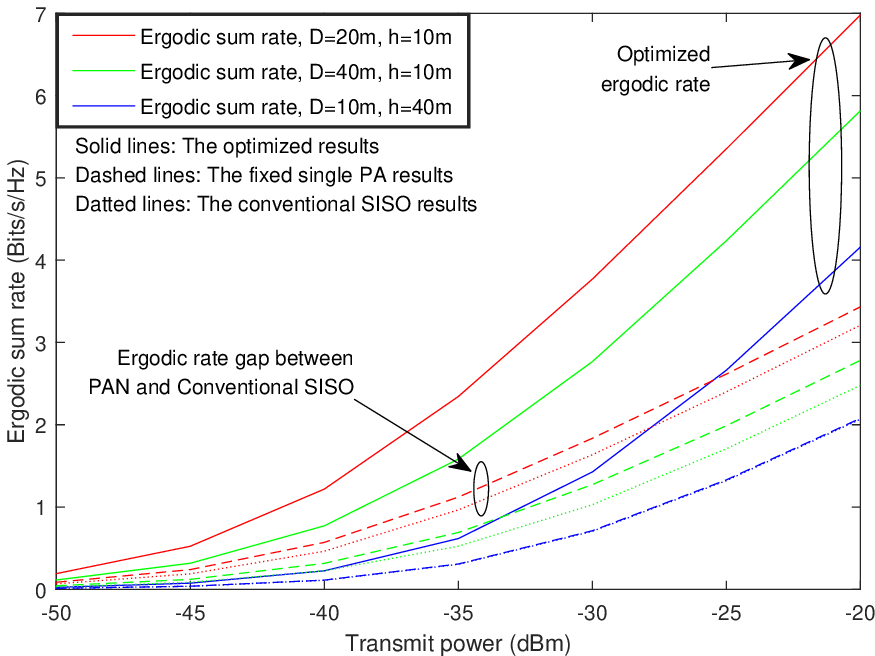}
\caption{Ergodic sum rate of the optimized SPMU scenario, fixed single SPMU scenario and conventional SISO scenario in UL transmission.}
\label{fig3+1_OMA_compare_opt}
\vspace{-0.1in}
\end{figure}

\emph{8) Ergodic sum rate comparison between optimized PASS and fixed PASS:}
We compare the ergodic sum rate between the optimized PA position and fixed PA position in SPMU scenario based on the results derived in~\eqref{Theo2+1 equa_large_Inte_single PA opt} and~\eqref{corro3 equati}, as illustrated in Fig.~\ref{fig3+1_OMA_compare_opt}. For reference, the performance of the conventional SISO is also included for the benchmark scheme. Solid curves correspond to the optimized PA placement results, while dashed and dotted curves represent the fixed PA and conventional SISO results, respectively. Note that in the conventional SISO setup, a single antenna is deployed at the AP located at the corner of cell edge, where the distance between antenna and user $i$ can be written as $\left| {{U_i} - L} \right| = \sqrt {x_{ui}^2 + y_{ui}^2 + {h^2}} $.
The optimization results indicate that optimizing the PA's position significantly enhances the ergodic sum rate compared to both the fixed PA and conventional SISO networks, emphasizing the critical role of future position optimization for PASS. In contrast, the fixed PA in SPMU scenario exhibits inferior performance due to the suboptimal placement of the PA. The conventional SISO system shows the lowest ergodic sum rate, highlighting the limitations of traditional fixed antenna systems in the indoor environments. These findings also validate our {\textbf{Remark~\ref{remark section II single PA OMA_opt position}}}. Moreover, the ergodic rate gap between the proposed optimized PA, fixed PA and conventional SISO networks is clearly observed in the figure particularly at higher transmit power regime, underscoring the performance benefits of the PASS—even in the fixed single-PA scenario.

\section{Conclusions}

In this article, we began by reviewed previous contributions related to flexible antennas, and we then proposed a novel PASS UL network. Three possible scenarios, MPSU, SPSU and SPMU scenarios, were considered.
For the MPSU and SPSU scenarios, the placement of PAs was optimized to minimize the distance to their corresponding users, thereby enhancing the channel gain. The ergodic rate of randomly located users was derived in closed-form. In order to glean deeper engineering insights, the asymptotic and approximated results were also derived. Then, to address the challenge of hardware complexity, a SPMU scenario was further investigated to serve all users, while suffering higher path loss.
For the future research, one potential direction is to design more dynamic PA placement strategies when the number of PAs is fewer than the number of users. Additionally, exploring the uplink performance of networks with multiple waveguides and PAs is another promising avenue for further enhancing system capacity and efficiency.

\numberwithin{equation}{section}
\section*{Appendix~A: Proof of Lemma~\ref{lemma_1_new_Optimized massive PAs location}} \label{Appendix1}
\renewcommand{\theequation}{A.\arabic{equation}}
\setcounter{equation}{0}

We first concentrate on the FA with $n>0$. For the signals received by each PA to be coherent, the differences in transmission distance between the $n$-th PA and its nearest PA must satisfy the condition: ${x_n} + {d_n} = {d_0} + n\lambda $, where $d_n$ represents the distance between the $i$-th user and the $n$-th PA in the FA. Note that this condition ensures that the phase shifts caused by the transmission distances are integer multiples of $\lambda$ for the FA, leading to coherent signal reception at the AP.

For the FA, based on the Pythagoras theorem that $x_n^2 + d_0^2 = d_n^2$, we have
\begin{equation}\label{appendix newA_FA_first equ}
{x_n} + \sqrt {x_n^2 + d_0^2}  = {d_0} + n\lambda .
\end{equation}

By some algebraic manipulations, we have
\begin{equation}\label{appendix newA_FA_results}
{x_n} = \frac{{n\lambda \left( {2{d_0} + n\lambda } \right)}}{{2\left( {{d_0} + n\lambda } \right)}},n \ge 0.
\end{equation}

On the contrary, for the NA where $n<0$, it is worth noting that $n$ is a negative number, and the following relation must hold if multiple signals are coherent:
\begin{equation}\label{appen_newA_NA_first relation}
{d_n} - n\lambda  = {d_0} + {x_n}.
\end{equation}

Based on the Pythagoras theorem again, and if ${d_0} > \left| n \right|\lambda $, we have
\begin{equation}\label{appen_newB_NA_result}
{x_n} = \frac{{ - n\lambda \left( {2{d_0} + n\lambda } \right)}}{{2\left( {{d_0} + n\lambda } \right)}},n < 0,
\end{equation}
and the proof is complete.

\numberwithin{equation}{section}
\section*{Appendix~B: Proof of Theorem~\ref{theorem New1 massive PAs ergodic rate}} \label{Appendix2}
\renewcommand{\theequation}{B.\arabic{equation}}
\setcounter{equation}{0}

In the FZ case, the following approximation can be realized:
\begin{equation}\label{appendix newB approximation of ln}
\ln \left( {\frac{{N\lambda }}{{d_0^{}}} + \sqrt {1 + {{\left( {\frac{{N\lambda }}{{d_0^{}}}} \right)}^2}} } \right) \to \frac{{N\lambda }}{{d_0^{}}}.
\end{equation}

Since the TDMA technique is adopted, thus the time resource blocks are equally allocated to each user. By substituting~\eqref{appendix newB approximation of ln} into~\eqref{SNR_massive PAs}, the data rate of the $i$-th user can be written as:
\begin{equation}\label{appendix newB data_rate}
R_{i,{\rm{mPAs}}} \approx \frac{1}{I}{\log _2}\left( {1 + \frac{{2N\gamma {P_i}}}{{d_0^2}}} \right).
\end{equation}

To calculate the closed-form expressions of the ergodic rate of the $i$-th user, and due to the fact that the distance between the $i$-th user and its nearest PA is randomly distributed in y-axis, we first reformulate the expectation as follows: \begin{equation}\label{AppenA_1}
\begin{aligned}
& \mathbb{ E} \left\{R_{i,{\rm{mPAs}}} \right\} =
 \underbrace {\frac{1}{ID}\int_0^D {{{\log }_2}\left( {y_{ui}^2 + {h^2} + 2N\gamma {P_i}} \right)} dy_{ui}^{}}_{{L_1}} \\
&- \underbrace {\frac{1}{ID}\int_0^D {{{\log }_2}\left( {y_{ui}^2 + {h^2}} \right)} dy_{ui}^{}}_{{L_2}}.
\end{aligned}
\end{equation}

Based on the insights from~\cite{Table_of_integrals}, we have
\begin{equation}\label{AppenA_2 integral temp}
\begin{aligned}
&\int_0^D {{{\log }_2}\left( {{x^2} + a} \right)dx}   = \frac{1}{{\ln 2}}\int_0^D {\ln \left( {{x^2} + a} \right)dx} \\
&  = D{\log _2}\left( {{D^2} + a} \right) + \frac{2}{{\ln 2}}\sqrt a \arctan \left( {\frac{D}{{\sqrt a }}} \right) - \frac{{2D}}{{\ln 2}}.
\end{aligned}
\end{equation}

By substituting~\eqref{AppenA_2 integral temp} into~\eqref{AppenA_1}, $L_1$ can be rewritten as:
\begin{equation}\label{AppenA_3_L1_expression}
\begin{aligned}
{L_1} & = {\log _2}\left( {D_{}^2 + {h^2} + 2N\gamma {P_i}} \right) - \frac{2}{{\ln 2}} \\
& + \frac{{2\sqrt {{h^2} + 2N\gamma {P_i}} }}{{D\ln 2}}\arctan \frac{D}{{\sqrt {{h^2} + 2N\gamma {P_i}} }}.
\end{aligned}
\end{equation}

Similarly, $L_2$ can be rewritten as:
\begin{equation}\label{AppenA_4_L2_express}
{L_2} = {\log _2}\left( {D_{}^2 + {h^2}} \right) - \frac{2}{{\ln 2}} + \frac{{2h}}{{D\ln 2}}\arctan \frac{D}{h}.
\end{equation}

Then, by some algebraic manipulations, the results in~\eqref{Theo1_equ_OMA_UL_Mul_PA} can be obtained, and the proof is complete.

\numberwithin{equation}{section}
\section*{Appendix~C: Proof of Corollary~\ref{Corro1_OMA_Mul_PA_approx}} \label{Appendix:Cs} \renewcommand{\theequation}{C.\arabic{equation}}
\setcounter{equation}{0}

In order to glean deeper insights, the inverse trigonometric function can be expanded by the Maclaurin's series, which is a special case of the Taylor series expansion as follows~\cite{Table_of_integrals}:
\begin{equation}\label{AppenB_expan}
\arctan x = \sum\limits_{k = 0}^\infty  {\frac{{{{\left( { - 1} \right)}^k}{x^{2k + 1}}}}{{2k + 1}}} ,
\end{equation}
when $x \ne \frac{\pi }{2} + k\pi $.
Thus, we have
\begin{equation}\label{Appen B_expression}
\begin{aligned}
& \sqrt {{h^2} + \gamma {P_i}} \arctan \frac{D}{{\sqrt {{h^2} + \gamma {P_i}} }} \\
& = \sum\limits_{k = 0}^\infty  {\frac{{{{\left( { - 1} \right)}^k}}}{{2k + 1}}} \frac{{{D^{2k + 1}}}}{{{{\left( {{h^2} + \gamma {P_i}} \right)}^k}}}.
\end{aligned}
\end{equation}

By some algebraic manipulations, the approximated achievable rate of the $i$-th user can be given by:
\begin{equation}\label{Appen B final expression}
\begin{aligned}
&{{\hat R}_i} = \frac{1}{I} {\log _2}\left( {1 + \frac{{\gamma {P_i}}}{{D_{}^2 + {h^2}}}} \right) \\
&+ \frac{2}{{I\ln 2}}\sum\limits_{k = 0}^\infty  {\frac{{{{\left( { - 1} \right)}^k}{D^{2k}}}}{{2k + 1}}} \left( {\frac{1}{{{{\left( {{h^2} + \gamma {P_i}} \right)}^k}}} - \frac{1}{{{h^{2k}}}}} \right),
\end{aligned}
\end{equation}
and the proof is complete.

\numberwithin{equation}{section}
\section*{Appendix~D: Proof of Corollary~\ref{Corro3_OMA_Sing_PA_approx}} \label{Appendix:Ds}
\renewcommand{\theequation}{D.\arabic{equation}}
\setcounter{equation}{0}

The proof starts by expressing $L_3$ in~\eqref{Theo2 equa_large_Inte} of user $i$ as follows:
\begin{equation}\label{AppenC_L3 first expression}
\begin{aligned}
&{L_3} =  \\
& \int_0^D {} \int_{ - \frac{D}{2}}^{\frac{D}{2}} {{{\log }_2}\left( {{{\left( {\frac{D}{2} - {x_{ui}}} \right)}^2} + y_{ui}^2 + {h^2} + \gamma {P_i}} \right)} d{x_{ui}}dy_{ui}^{}.
\end{aligned}
\end{equation}

We starts from the integral of variable $x_{ui}$, thus we have
\begin{equation}\label{AppenC_L3_integral x}
\begin{aligned}
{L_3}&= \underbrace {\int_0^D {} D{{\log }_2}\left( {4{D^2} + 4y_{ui}^2 + 4{h^2} + 4\gamma {P_i}} \right)}_{{L_{3,1}}} \\
&+ \frac{{2\sqrt {y_{ui}^2 + {h^2} + \gamma {P_i}} }}{{\ln 2}}\arctan \frac{D}{{\sqrt {y_{ui}^2 + {h^2} + \gamma {P_i}} }} \\
&- 2D - \frac{{2D}}{{\ln 2}}dy_{ui}^{}.
\end{aligned}
\end{equation}

We first calculus $L_{3,1}$ in~\eqref{AppenC_L3_integral x}, which can be rewritten as:
\begin{equation}\label{AppenC L31_expression_exact}
\begin{aligned}
&{L_{3,1}} = {D^2}{\log _2}\left( {4\left( {2D_{}^2 + {h^2} + \gamma {P_i}} \right)} \right) \\
&+ \frac{{D\sqrt {4{D^2} + {h^2} + \gamma {P_i}} }}{{\ln 2}}\arctan \frac{D}{{\sqrt {4{D^2} + {h^2} + \gamma {P_i}} }} - \frac{{2{D^2}}}{{\ln 2}}.
\end{aligned}
\end{equation}

In the high-SNR regime, by adopting $\arctan x \to x$, ~\eqref{AppenC L31_expression_exact} can be further transformed into:
\begin{equation}\label{AppenC_L31_approximateion}
{L_{3,1}} = {D^2}{\log _2}\left( {4\left( {2D_{}^2 + {h^2} + \gamma {P_i}} \right)} \right) - \frac{{{D^2}}}{{\ln 2}}.
\end{equation}

Similarly, the rest parts in~\eqref{AppenC L31_expression_exact} can be calculated as:
\begin{equation}\label{AppenC_L3_additionpart}
\int_0^D { - 2Dd} y_{ui}^{} =  - 2{D^2}.
\end{equation}

Thus, we have
\begin{equation}\label{AppenC_L3_approximate_result}
\begin{aligned}
{L_3} = {D^2}{\log _2}\left( {4\left( {2D_{}^2 + {h^2} + \gamma {P_i}} \right)} \right) - \frac{{{D^2}}}{{\ln 2}} - 2{D^2}.
\end{aligned}
\end{equation}

We then use the similar steps from~\eqref{AppenC_L3 first expression} to~\eqref{AppenC_L3_approximate_result}, and $L_4$ can be obtained as follows:
\begin{equation}\label{AppenC_L4_exact}
\begin{aligned}
{L_4} &= \underbrace {\int_0^D {} D{{\log }_2}\left( {4{D^2} + 4y_{ui}^2 + 4{h^2}} \right)}_{{L_{4,1}}} \\
&+ \underbrace {\frac{{2\sqrt {y_{ui}^2 + {h^2}} }}{{\ln 2}}\arctan \frac{D}{{\sqrt {y_{ui}^2 + {h^2}} }}}_{{L_{4,2}}} \\
&- 2D - \frac{{2D}}{{\ln 2}} dy_{ui}^{}.
\end{aligned}
\end{equation}

$L_{4,1}$ can be simply derived as follows:
\begin{equation}\label{AppenC_L41}
\begin{aligned}
{L_{4,1}} &= {D^2}{\log _2}\left( {4\left( {2D_{}^2 + {h^2}} \right)} \right) \\
&+ \frac{{D\sqrt {4{D^2} + {h^2}} }}{{\ln 2}}\arctan \frac{D}{{\sqrt {4{D^2} + {h^2}} }} - \frac{{{D^2}}}{{\ln 2}}.
\end{aligned}
\end{equation}

Regarding $L_{4,2}$, it is assumed that $h \gg D$, the approximation can be then applied, and is expressed as:
\begin{equation}\label{AppenC_L42}
{L_{4,2}} \approx \int_0^D {} \frac{{2D}}{{\ln 2}}dy_{ui}^{} = \frac{{2{D^2}}}{{\ln 2}}.
\end{equation}

Thus, we have
\begin{equation}\label{AppenC L4}
\begin{aligned}
{L_4} &\approx {D^2}{\log _2}\left( {4\left( {2D_{}^2 + {h^2}} \right)} \right) \\
&+ \frac{{D\sqrt {4{D^2} + {h^2}} }}{{\ln 2}}\arctan \frac{D}{{\sqrt {4{D^2} + {h^2}} }} - 2{D^2} - \frac{{{D^2}}}{{\ln 2}}.
\end{aligned}
\end{equation}

By some mathematical manipulations, we have the results in~\eqref{corro3 equati}, and the proof is complete.

\numberwithin{equation}{section}
\section*{Appendix~E: Proof of Proposition~\ref{propo1_OMA compare}} \label{Appendix:Es}
\renewcommand{\theequation}{E.\arabic{equation}}
\setcounter{equation}{0}

In order to compare the MPSU scenario and SPSU scenario in UL transmission, the ergodic rate difference between multiple PA scenario and single PA scenario can be reformulated as:
\begin{equation}\label{AppenD_first expre}
R_{i,diff} = \bar R_{i,{\rm{PA}}s} - \hat R_{i,{\rm{A1}}}.
\end{equation}

It is obvious that ${D_{}^2 + {h^2}}  \le {2D_{}^2 + {h^2}}$, thus we have
\begin{equation}\label{compare log part}
{\log _2}\left( {1 + \frac{{\gamma {P_i}}}{{D_{}^2 + {h^2}}}} \right) - {\log _2}\left( {1 + \frac{{\gamma {P_i}}}{{2D_{}^2 + {h^2}}}} \right) \ge 0,
\end{equation}
which approaches 0 when $D=0$.

We further examine the rest part when $D=0$, we have $2 - \frac{{2h}}{D}\arctan \frac{D}{h} \to 0$. Clearly, $\frac{{\sqrt {4{D^2} + {h^2}} }}{D}\arctan \frac{D}{{\sqrt {4{D^2} + {h^2}} }} \ge 0$ is always established.

On the contrary, when $D \to \infty$, we have $\arctan \frac{D}{h} \to \frac{\pi }{2}$, resulting in $\frac{{2h}}{D}\arctan \frac{D}{h} \to \frac{{\pi h}}{D} \to 0$. Thus, the comparison can be explored as follows:
\begin{equation}\label{AppenD_final expre}
 \bar R_{i,{\rm{PA}}s} > \hat R_{i,{\rm{A1}}} ,
\end{equation}
and the proof is complete.

\numberwithin{equation}{section}
\section*{Appendix~F: Proof of Theorem~\ref{Theo2+1:single PA_OMA_opt}} \label{Appendix:Fs}
\renewcommand{\theequation}{F.\arabic{equation}}
\setcounter{equation}{0}

The proof starts by expression the achievable ergodic rate of two randomly deployed users by:
\begin{equation}\label{Appen new E OMA opt user1 rate}
R_{1,OPT} = \frac{1}{2}{\log _2}\left( {1 + \frac{{\gamma {P_1}}}{{{{\left( {x - {x_{u1}}} \right)}^2} + y_{u1}^2 + {h^2}}}} \right),
\end{equation}
and
\begin{equation}\label{Appen new E OMA opt user2 rate}
R_{2,OPT} = \frac{1}{2}{\log _2}\left( {1 + \frac{{\gamma {P_2}}}{{{{\left( {x - {x_{u2}}} \right)}^2} + y_{u2}^2 + {h^2}}}} \right).
\end{equation}

Then, the optimization problem can be given by:
\begin{equation}\label{Appen new E OMA opt sum rate}
\begin{array}{l}
\max~ R_{1,OPT} + R_{2,OPT}\\
s.t.~~~ {x_{u1}} \le x \le {x_{u2}},
\end{array}
\end{equation}
where it is evident that the optimal solution for $x$ must lie between ${x_{u1}}$ and ${x_{u2}}$.

In the indoor scenario, we only consider the high-SNR approach, where $\frac{{{P_i}}}{{{\sigma ^2}}} \to \infty $, thus we have
\begin{equation}\label{Append new E sum rate_expres 2}
\begin{aligned}
& R_{OPT}  = R_{1,OPT} + R_{2,OPT} \\
&  \approx  \frac{1}{2}{\log _2}\left( {\frac{{\gamma {P_1}}}{{{{\left( {x - {x_{u1}}} \right)}^2} + y_{u1}^2 + {h^2}}}\frac{{\gamma {P_2}}}{{{{\left( {x - {x_{u2}}} \right)}^2} + y_{u2}^2 + {h^2}}}} \right)\\
& = \frac{1}{2}{\log _2}\left( {{\gamma ^2}{P_1}{P_2}} \right) - \frac{1}{2}{\log _2}\left( {{{\left( {x - {x_{u1}}} \right)}^2} + y_{u1}^2 + {h^2}} \right) \\
& - \frac{1}{2}{\log _2}\left( {{{\left( {x - {x_{u2}}} \right)}^2} + y_{u2}^2 + {h^2}} \right).
\end{aligned}
\end{equation}

We then derive the derivative of~\eqref{Append new E sum rate_expres 2} with respect to $x$, which can be written as:
\begin{equation}\label{Appen new E derivative}
\begin{aligned}
&\frac{{\partial R_{OPT}}}{{\partial x}} = \\
& \frac{{ - 1}}{{\ln (2)}}\left( {\frac{{\left( {x - {x_{u1}}} \right)}}{{{{\left( {x - {x_{u1}}} \right)}^2} + y_{u1}^2 + {h^2}}} + \frac{{\left( {x - {x_{u2}}} \right)}}{{{{\left( {x - {x_{u2}}} \right)}^2} + y_{u2}^2 + {h^2}}}} \right).
\end{aligned}
\end{equation}

Obviously, the optimized position of PA can be find by setting~\eqref{Appen new E derivative} to 0, thus we have:
\begin{equation}\label{Appen new E final}
\frac{{\left( {x - {x_{u1}}} \right)}}{{{{\left( {x - {x_{u1}}} \right)}^2} + y_{u1}^2 + {h^2}}} = \frac{{\left( {{x_{u2}} - x} \right)}}{{{{\left( {x - {x_{u2}}} \right)}^2} + y_{u2}^2 + {h^2}}}.
\end{equation}

After some algebraic manipulations, we have the results in~\eqref{Theo2+1 equa_large_Inte_single PA opt}, and the proof is complete.

\bibliographystyle{IEEEtran}
\bibliography{IEEEabrv,CAPA_NOMA}

\end{document}